\documentclass[12pt,preprint]{emulateapj}

\begin{document}

\title{Unusually Wide Binaries: Are They Wide or Unusual?}

\author{Adam L. Kraus}
\email{alk@astro.caltech.edu}
\author{Lynne A. Hillenbrand}
\email{lah@astro.caltech.edu}

\affil{California Institute of Technology, Department of Astrophysics, MC 105-24, Pasadena, CA 91125}

\begin{abstract}

We describe an astrometric and spectroscopic campaign to confirm the youth 
and association of a complete sample of candidate wide companions in 
Taurus and Upper Sco. Our survey found fifteen new binary systems (3 in 
Taurus and 12 in Upper Sco) with separations of 3-30\arcsec\, (500-5000 
AU) among all of the known members with masses of 2.5-0.012 $M_{\sun}$. 
The total sample of 49 wide systems in these two regions conforms to only 
some expectations from field multiplicity surveys. Higher-mass stars have 
a higher frequency of wide binary companions, and there is a marked 
paucity of wide binary systems near the substellar regime. However, the 
separation distribution appears to be log-flat, rather than declining as 
in the field, and the mass ratio distribution is more biased toward 
similar-mass companions than the IMF or the field G dwarf distribution. 
The maximum separation also shows no evidence of a limit at $\la$5000 AU 
until the abrupt cessation of any wide binary formation at system masses 
of $\sim$0.3 $M_{\sun}$. We attribute this result to the post-natal 
dynamical sculpting that occurs for most field systems; our binary systems 
will escape to the field intact, but most field stars are formed in denser 
clusters and do not. In summary, only wide binary systems with total 
masses $\la$0.3 $M_{\sun}$ appear to be ``unusually wide''.

\end{abstract}

\keywords{stars:binaries:visual, stars:formation, 
stars:pre-main-sequence}

\section{Introduction}

The frequency and properties of multiple star systems are important 
diagnostics for placing constraints on star formation processes. This 
motivation has prompted numerous attempts to characterize the properties 
of nearby binary systems in the field. These surveys (e.g. Duquennoy \& 
Mayor 1991; Fischer \& Marcy 1992; Close et al. 2003; Bouy et al. 2003; 
Burgasser et al. 2003) have found that binary frequencies and properties 
are very strongly dependent on mass. Solar-mass stars have high binary 
frequencies ($\ga$60\%) and maximum separations of up to $\sim$10$^4$ AU. 
By contrast, M dwarfs have moderately high binary frequencies (30-40\%) 
and few binary companions with separations of more than $\sim$500 AU, 
while brown dwarfs have low binary frequencies ($\sim$15\% for all 
companions with separations $\ga$2-4 AU) and few companions with 
separations $>$20 AU.

The mass-dependent decline in the maximum observed binary 
separation (or binding energy) has been described by Reid et al. 
(2001) and Burgasser et al. (2003) with an empirical function that 
is exponential at high masses and quadratic at low masses. The 
mechanism that produces the mass dependence is currently unknown. 
Simulations show that the empirical limit is not a result of 
dynamical evolution in the field (e.g. Burgasser et al. 2003; 
Weinberg et al. 1987) since the rates of binary disruption (due to 
single stellar encounters with small impact parameters) and 
evolution in the separation distribution (due to many encounters 
at large impact parameters) are far too low. This suggests that 
the limit must be set early in stellar lifetimes, either as a 
result of the binary formation process or during early dynamical 
evolution in relatively crowded natal environments.

Studies of nearby young stellar associations have identified 
several candidate systems which might be unusually wide binaries 
(Chauvin et al. 2004; Caballero et al. 2006; Jayawardhana \& 
Ivanov 2006; Luhman et al. 2006a, 2007; Close et al. 2007; Kraus 
\& Hillenbrand 2007b). However, there are several factors that 
must be considered when interpreting these discoveries. Most were 
identified serendipitously and not as part of a survey, so the 
actual frequency of these candidates is not well constrained. 
Further, several of these systems do not seem to be unusual in 
comparison to field systems of similar mass. Finally, many of 
these systems have not been surveyed at high angular resolution, 
so they could be hierarchical multiples with higher total masses.

We began addressing these problems by using archival 2MASS data to 
systematically search for candidate wide binary systems among all 
of the known members of three nearby young associations (Upper 
Sco, Taurus-Auriga, and Chamaeleon-I; Kraus \& Hillenbrand 2007a, 
hereafter KH07a). Our results broadly agreed with the standard 
paradigm; there is a significant deficit of wide systems among 
very low-mass stars and brown dwarfs as compared to their more 
massive brethren. However, we also found that most of these wide 
systems were concentrated in the very sparsest T associations, 
Taurus and Cham-I. Upper Sco is not significantly more dense than 
either of these associations, so it is unclear why it might have 
such a meager wide binary population. We also found a few 
candidate systems which appeared to be unusually wide for their 
mass. However, photometric criteria alone are not sufficient to 
reject all background stars.

In this paper, we describe our astrometric and spectroscopic 
followup campaign to confirm or reject the youth and association 
of our new sample of candidate binary companions. In Section 2, we 
describe the compilation of our sample of candidate wide binary 
systems. In Section 3, we describe the observations and analysis 
conducted for our survey, and in Section 4, we evaluate this 
evidence in order to distinguish association members from field 
stars. Finally, in Section 5, we describe the mass-dependent 
binary frequency, mass ratio distribution, and separation 
distribution of these systems, plus we examine the criteria that 
might define an ``unusually wide'' binary system.

\section{Sample}

We drew the sample from our previous companion search (KH07a), which 
used 2MASS photometry to identify candidate companions to members of 
Taurus, Upper Sco, and Cham-I. The survey used PSF-fitting photometry of 
the 2MASS atlas images to identify close (1-5\arcsec) companions and 
archival data from the 2MASS Point Source Catalog to identify 
well-resolved ($\ga$5\arcsec) companions. For this study, we do not 
include any of the candidates in Cham-I or the southern subgroup of 
Upper Sco (USco-B) since our observations were all conducted from 
northern sites. We consider every candidate in the other two 
associations with a separation of $>$3\arcsec\, (out to a limit of 
30\arcsec) and a flux ratio of $\Delta$$K\la$3 (corresponding to mass 
ratios $q\ga$0.1). We also considered all 14 candidates in Taurus with 
larger flux ratios, yielding a complete sample down to the 10$\sigma$ 
flux limit of 2MASS ($K=14.3$); we were not able to gather sufficient 
information to consider one of the 3 candidates with large flux ratios in 
Upper Sco.

We list all of the previously-unconfirmed candidate companions in our 
sample in Table 1. Some of the sources in our sample have been identified 
previously in the literature as either field stars or association members 
based on a wide variety of characteristics: proper motions, the presence 
of a disk, low surface gravity, or the presence of lithium. We summarize 
these identifications in Tables 2 and 3, respectively. Table 3 also 
includes all of the systems we identified in a similar compilation in 
KH07a.

Finally, in Tables 1-3 we have compiled updated spectral types for all 
members of our sample. Our original survey used the spectral types 
assigned in the discovery survey or in compilation papers (e.g. Kenyon \& 
Hartmann 1995), but a significant number of system components have had 
more precise spectral type estimates published since their discovery. 
Unless otherwise noted, the masses were estimated using the methods 
described in Section 3.4. In hierarchical multiple systems where 
components are themselves known to be multiple from previous AO, speckle, 
or RV surveys, we have noted the known or estimated spectral type of each, 
and report the corresponding known or estimated system mass. We also have 
updated the spectral types and multiplicity (and therefore the masses) for 
all sample members that do not have wide companions, so the analysis in 
Section 5 is performed with a uniform sample.

\begin{deluxetable*}{llrrrll}
\tabletypesize{\tiny}
\tablewidth{0pt}
\tablecaption{Candidate Wide Companions in Taurus and Upper Sco}
\tablehead{\colhead{Known Member} & \colhead{Candidate Companion} &
\colhead{Sep} & \colhead{PA} & \colhead{$\Delta$$K$} & 
\colhead{$SpT_{known}$\tablenotemark{a}} & \colhead{Ref} 
\\
\colhead{} & \colhead{} & \colhead{(\arcsec)} & \colhead{(deg)} & 
\colhead{(mag)}
}
\startdata
\multicolumn{7}{c}{Taurus}\\
2M04414489+2301513&2M04414565+2301580&12.37&57.3&-3.31&M8.25&1\\
2M04080782+2807280&2M04080771+2807373&9.43&351.1&-1.96&M3.75&1\\
LkCa 15&2M04391795+2221310&27.62&4.6&-1.14&K5&4\\
FW Tau&2M04292887+2616483&12.22&246.7&0.03&M5.5+?&3\\
GM Aur&2M04551015+3021333&28.31&202.2&0.28&K7&4\\
2M04161885+2752155&2M04161754+2751534&28.04&218.2&0.60&M6.25&1\\
CFHT-Tau-7&2M04321713+2421556&21.76&207.2&0.82&M5.75&1\\
HBC 427&2M04560252+3020503&14.90&154.0&0.89&K5+?&5\\
I04158+2805&2M04185906+2812456&25.34&28.9&0.98&M5.25&1\\
JH 112&2M04324938+2253082&6.56&34.3&1.03&K7&7\\
V710 Tau AB&2M04315968+1821305&27.97&105.7&1.39&M0.5+M2&9\\
CFHT-Tau-21&2M04221757+2654364&23.31&152.1&2.06&M2&1\\
V410 X-ray1&2M04175109+2829157&27.95&137.4&2.66&M4&8\\
2M04213460+2701388&2M04213331+2701375&17.18&265.7&2.70&M5.5&2\\
I04385+2550&2M04413842+2556448&18.94&343.3&3.03&M0.5&6\\
DO Tau&2M04382889+2611178&28.75&8.4&3.28&M0&7\\
CFHT 4&2M04394921+2601479&24.40&72.9&3.56&M7&18\\
I04216+2603&2M04244376+2610398&27.96&337.0&3.66&M0&7\\
V410 X-ray 5a&2M04190271+2822421&13.27&47.7&3.67&M5.5&21\\
V410 X-ray 6&2M04190223+2820039&26.49&34.4&4.22&M5.5&21\\
MHO-Tau-2&2M04142440+2805596&26.32&269.9&4.32&M2.5+M2.5&20\\
V410 X-ray 2&2M04183574+2830254&17.72&105.6&4.47&M0&22\\
IS Tau&2M04333746+2609550&10.85&57.4&4.82&M0+M3.5&19\\
CoKu Tau/3&2M04354076+2411211&12.60&349.2&4.97&M1&7\\
FM Tau&2M04141556+2812484&26.21&91.7&4.98&M0&7\\
LkCa 4&2M04162839+2807278&8.86&154.6&5.25&K7&7\\
IS Tau&2M04333467+2609447&28.73&261.1&5.64&M0+M3.5&19\\
FO Tau&2M04144741+2812219&26.19&250.8&5.98&M3.5+M3.5&19\\
DG Tau&2M04270370+2606067&16.43&234.3&6.71&K2&6\\
\multicolumn{7}{c}{Upper Sco}\\
SCH160758.50-203948.90&2M16075796-2040087&21.52&200.7&-4.78&M6&13\\
USco80&2M15583621-2348018&12.27&15.2&-1.89&M4&17\\
DENIS162041.5-242549.0&2M16204196-2426149&26.73&164.5&-1.28&M7.5&10\\
SCH161511.15-242015.56&2M16151239-2420091&17.96&69.8&-1.04&M6&13\\
UScoJ160700.1-203309&2M16065937-2033047&11.65&293.1&-0.40&M2&16\\
SCH161825.01-233810.68&2M16182365-2338268&24.73&229.1&-0.20&M5&13\\
SCH162135.91-235503.41&2M16213638-2355283&25.65&165.3&-0.19&M6&13\\
ScoPMS048&ScoPMS 048 B&3.05&192.1&0.25&K2+M4&15\\
SCH160758.50-203948.90&2M16075693-2039424&22.94&285.5&1.39&M6&13\\
RXJ 1555.8-2512&2M15554839-2512174&8.91&318.4&1.71&G3&12\\
RXJ 1558.8-2512&2M15585415-2512407&11.35&130.1&1.88&M1&12\\
GSC 06213-01459&GSC 06213-01459 B&3.18&305.5&2.14&K5&11\\
UScoJ160936.5-184800&2M16093658-1847409&19.97&2.2&2.22&M3&16\\
ScoPMS042b&2M16102177-1904021&4.58&6.8&2.31&M3&14\\
RXJ 1602.8-2401B&2M16025116-2401502&7.22&352.9&2.69&K4&12\\
UScoJ160245.4-193037&2M16024735-1930294&28.19&72.9&2.74&M5&16\\
GSC 06784-00997&2M16101888-2502325&4.81&240.4&2.90&M1&11\\
GSC 06785-00476&2M15410726-2656254&6.30&82.6&3.04&G7&11\\
UScoJ161031.9-191305&2M16103232-1913085&5.71&114.0&3.74&K7&16\\
RXJ 1555.8-2512&2M15554788-2512172&14.61&298.1&4.24&G3&12\\
GSC 06784-00039&2M16084438-2602139&13.53&77.5&5.12&G7&11\\
\enddata
\tablenotetext{a}{Entries with multiple spectral types denote components 
which are themselves known to be multiple; if the spectral type for a 
component has not been measured, it is listed as ``?''.}
 \tablecomments{The astrometry and photometry for each candidate system 
have been adopted from our re-reduction of the 2MASS atlas images (KH07a). 
References: (1) Luhman (2006a), (2) Luhman (2004), (3) White \& Ghez 
(2001), (4) Simon et al. (2000), (5) Steffen et al. (2001), (6) White \& 
Hillenbrand (2004), (7) Kenyon \& Hartmann (1995), (8) Strom \& Strom 
(1994), (9) Hartigan et al. (1994), (10) Mart\'in et al. (2004), (11) 
Preibisch et al. (1998), (12) Kunkel (1999), (13) Slesnick et al. (2006a), 
(14) Walter et al. (1994), (15) Prato et al. (2002a), (16) Preibisch et 
al. (2002), (17) Ardila et al. (2000), (18) Mart\'in et al. (2001), (19) 
Hartigan \& Kenyon (2003), (20) Briceno et al. (1998), (21) Luhman (1999), 
(22) Luhman \& Rieke (1998).}
 \end{deluxetable*}

\begin{deluxetable*}{llrrrll}
\tabletypesize{\tiny}
\tablewidth{0pt}
\tablecaption{Previously Confirmed Field Stars}
\tablehead{\colhead{Known Member} & \colhead{Field Star} &
\colhead{Sep} & \colhead{PA} & \colhead{$\Delta$$K$} & \colhead{Evidence} & 
\colhead{Ref}
\\
\colhead{} & \colhead{} & \colhead{(\arcsec)} & \colhead{(deg)} & 
\colhead{(mag)}
}
\startdata
IP Tau&NLTT 13195&15.75&55.7&5.08&Proper Motion&Salim \& Gould (2003)\\
V410 Anon 20&V410 Anon 21&22.71&115.3&0.62&Early SpT&Luhman (2000)\\
USco160428.4-190441&GSC06208-00611&24.15&134.3&0.51&Lithium&Preibisch et al. (1998)\\
USco161039.5-191652&SIPS1610-1917&14.95&183.2&1.98&Proper Motion&Deacon \& Hambly (2007)\tablenotemark{a}\\
\enddata
\tablecomments{The astrometry and photometry for each pair of stars have 
been adopted from our re-reduction of the 2MASS atlas images 
(KH07a).}
\tablenotetext{a}{Deacon \& Hambly (2007) identified SIPS1610-1917 as 
USco161039.5-191652, but inspection of the original photographic plates 
shows that SIPS1610-1917 is the candidate companion that we identified in 
KH07a (2M16103950-1917073). Its high proper motion demonstrates that 
it is a field star, not a bound companion.}
\end{deluxetable*}

\begin{deluxetable*}{llrrrlll}
\tabletypesize{\tiny}
\tablewidth{0pt}
\tablecaption{Previously Confirmed companions}
\tablehead{\colhead{Primary} & \colhead{Secondary} & \colhead{Sep} & 
\colhead{PA} & \colhead{$\Delta$$K$} & 
\colhead{$SpT_{prim}$\tablenotemark{a}} & 
\colhead{$SpT_{sec}$\tablenotemark{a}} & \colhead{Refs} 
\\
\colhead{} & \colhead{} & \colhead{(\arcsec)} & \colhead{(deg)} & 
\colhead{(mag)}
}
\startdata
\multicolumn{8}{c}{Taurus}\\
2M04554757+3028077&2M04554801+3028050&6.31&115.7&2.18&M4.75&M5.5&1\\
DH Tau&DI Tau&15.23&126.0&0.21&M0+M7.5&M0+?&2, 3, 4\\
FS Tau&Haro 6-5B&19.88&275.8&3.57&M0+M3.5&K5&5, 6\\
FV Tau&FV Tau/c&12.29&105.7&1.43&K5+cont&M2.5+M3.5&2, 3, 5\\
FZ Tau&FY Tau&17.17&250.5&0.70&K7&M0&2\\
GG Tau Aab&GG Tau Bab&10.38&185.1&2.61&K7+M0.5&M5.5+M7.5&7\\
GK Tau&GI Tau&13.14&328.4&0.42&K7+cont&K5&2, 8\\
HBC 352&HBC 353&8.97&70.8&0.28&G5&K3&2\\
HBC 355&HBC 354&6.31&298.3&0.91&K2&K2&2\\
HN Tau A&HN Tau B&3.10&18.7&3.19&K5&M4.5&8, 9\\
HP Tau-G2&HP Tau&21.30&296.9&0.40&G0&K3&2, 10\\
HP Tau-G2&HP Tau-G3&10.09&243.4&1.57&G0&K7+?&2, 10\\
HV Tau AB&HV Tau C&3.76&43.9&4.35&M2+?&K6&6, 11\\
J1-4872 Aab&J1-4872 Bab&3.38&232.9&0.69&M0+M0&M1+M1&8\\
LkHa332-G1&LkHa332-G2&25.88&254.5&0.28&M1+?&M0.5+M2.5&2, 5, 12\\
MHO-Tau-1&MHO-Tau-2&3.93&153.9&0.01&M2.5&M2.5&13\\
UX Tau AC&UX Tau Bab&5.856&269.7&2.22&K2+M3&M2+?&8\\
UZ Tau Aab&UZ Tau Bab&3.56&273.5&0.24&M1+?&M2+M3&2, 5, 14, 15\\
V710 Tau A&V710 Tau B&3.03&178.5&-0.13&M0.5&M2&9\\
V773 Tau&2M04141188+2811535&23.38&215.9&5.43&K2+K5+M0.5+?&M6.25&1, 15\\
V807 Tau&GH Tau&21.77&195.2&0.83&K5+M2+?&M2+M2&5, 25\\
V928 Tau&CFHT-Tau-7&18.25&228.2&2.27&M0.5+?&M5.75&2, 16, 17\\
V955 Tau&LkHa332-G2&10.51&35.3&0.01&K7+M2.5&M0.5+M2.5&2, 5, 12\\
XZ Tau&HL Tau&23.31&271.2&0.12&M2+M3.5&K5&2, 5\\
\multicolumn{8}{c}{Upper Sco}\\
RXJ1558.1-2405A&RXJ1558.1-2405B&18.15&254.4&2.10&K4+?&M5+?&18, 19\\
RXJ1604.3-2130A&RXJ1604.3-2130B&16.22&215.9&0.92&K2&M2+?&18, 19\\
ScoPMS 052&RXJ1612.6-1859&19.06&269.5&1.62&K0+M2&M1&20, 21\\
UScoJ160428.4-190441&UScoJ160428.0-19434&9.77&321.3&1.73&M3+?&M4&22, 23\\
UScoJ160611.9-193532 A&UScoJ160611.9-193532 B&10.78&226.5&0.76&M5+M5&M5&22, 24\\
UScoJ160707.7-192715&UScoJ160708.7-192733&23.45&140.4&1.37&M2+?&M4&22\\
UScoJ160822.4-193004&UScoJ160823.2-193001&13.47&71.4&0.41&M1&M0&22\\
UScoJ160900.7-190852&UScoJ160900.0-190836&18.92&326.5&1.81&M0&M5&22\\
UScoJ161010.4-194539&UScoJ161011.0-194603&25.59&160.8&0.97&M3&M5&22\\
\enddata
\tablenotetext{a}{Entries with multiple spectral types denote components 
which are themselves known to be multiple; if the spectral type for a 
component has not been measured, it is listed as ``?''. Sources 
labeled "cont" only exhibit continuum emission from accretion and disk 
emission, with no recognizeable spectral features.}
\tablecomments{The astrometry and photometry for each candidate system have 
been adopted from our re-reduction of the 2MASS atlas images (KH07a). 
References: (1) Luhman (2004), (2) Kenyon \& Hartmann (1995), (3) Ghez 
et al. (1993), (4) Itoh et al. (2005), (5) Hartigan \& Kenyon (2003), (6) White 
\& Hillenbrand (2004), (7) White et al. (1999), (8) Duchene et al. (1999), (9) 
Hartigan et al. (1994), (10) Simon et al. (1995), (11) Stapelfeldt et al. 
(2003), (12) White \& Ghez (2001), (13) Briceno et al. (1998), (14) Prato et 
al. (2002b), (15) Correia et al. (2006), (15) Boden et al. (2007), (16) Simon 
et al. (1996), (17) Luhman (2006), (18) Kunkel (1999), (19) K\"ohler et al. 
(2000), (20) Walter et al. (1994), (21) Prato (2007), (22) Preibisch et 
al. (2002), (23) Kraus et al. (2008), (24) Kraus \& Hillenbrand 
(2007b), (25) Schaefer et al. (2006).}
\end{deluxetable*}

\section{Observations and Analysis}

\subsection{Optical Spectroscopy}

We obtained intermediate-resolution optical spectra for 14 Taurus 
candidates and 8 Upper Sco candidates that were wide enough to be easily 
resolved and optically bright enough to be observed with short ($\le$10 
min) exposures. These spectra were measured with the Double Spectrograph 
(Oke \& Gunn 1982) on the Hale 5m telescope at Palomar Observatory in 
December 2006 and May 2007. The spectra presented here were obtained with 
the red channel using a 316 l/mm grating and a 2.0\arcsec\, slit, yielding 
a spectral resolution of $R\sim$1250 over a wavelength range of 6400-8800 
angstroms. Wavelength calibration was achieved by observing a standard 
lamp after each science target, and flux normalization was achieved by 
periodic observation of spectrophotometric standard stars from the 
compilation by Massey et al. (1988). We summarize all of the observations 
in Table 4.

The spectra were processed using standard IRAF\footnote{IRAF is 
distributed by the National Optical Astronomy Observatories, which are 
operated by the Association of Universities for Research in Astronomy, 
Inc., under cooperative agreement with the National Science Foundation.} 
tasks; we used the IRAF task SPLOT to measure equivalent widths of 
spectral lines. Several of the fainter candidates have very noisy spectra 
because we recognized from short preliminary exposures that they were 
heavily reddened background stars and not late-type association members; 
given their brightness and color, these candidates would possess deep TiO 
bands if they were members.

\begin{deluxetable}{lcr}
\tabletypesize{\scriptsize}
\tablewidth{0pt}
\tablecaption{Spectroscopic Observations}
\tablehead{\colhead{Candidate} & \colhead{Instrument} & 
\colhead{$t_{int}$}
\\
\colhead{Companion} & \colhead{} & \colhead{(sec)}
}
\startdata
2M04080771+2807373&DBSP&300\\
2M04161754+2751534&DBSP&300\\
2M04213331+2701375&DBSP&600\\
2M04414565+2301580&DBSP&300\\
2M04394921+2601479&NIRSPEC&300\\
2M04221757+2654364&DBSP&300\\
2M04321713+2421556&DBSP&300\\
2M04354076+2411211&NIRSPEC&300\\
2M04382889+2611178&NIRSPEC&300\\
2M04141556+2812484&NIRSPEC&300\\
2M04292887+2616483&DBSP&300\\
2M04551015+3021333&DBSP&30\\
2M04560252+3020503&DBSP&240\\
2M04185906+2812456&DBSP&300\\
2M04244376+2610398&NIRSPEC&300\\
2M04333746+2609550&NIRSPEC&300\\
2M04324938+2253082&DBSP&600\\
2M04391795+2221310&DBSP&60\\
2M04162839+2807278&NIRSPEC&300\\
2M04142440+2805596&NIRSPEC&300\\
2M04190271+2822421&NIRSPEC&300\\
2M04175109+2829157&DBSP&300\\
2M04183574+2830254&NIRSPEC&300\\
2M04315968+1821305&DBSP&600\\
2M04190223+2820039&NIRSPEC&300\\
2M16204196-2426149&DBSP&300\\
2M15554839-2512174&DBSP&300\\
2M16075796-2040087&DBSP&60\\
2M16151239-2420091&DBSP&300\\
2M16182365-2338268&DBSP&300\\
2M16213638-2355283&DBSP&300\\
2M15583621-2348018&DBSP&180\\
2M16065937-2033047&DBSP&60\\
\enddata
\end{deluxetable}

\subsection{Near-Infrared Spectroscopy}

We obtained intermediate-resolution near-infrared spectra for 11 of our 
Taurus candidates that were too faint and red for optical spectroscopy. 
These spectra were obtained using NIRSPEC on the Keck-II 10m telescope on 
JD 2454398 with the NIRSPEC-7 (K) filter using the low-resolution grating 
and a 0.76\arcsec\, slit. The corresponding spectral resolution is 
$R\sim$1500 spanning 1.95-2.37 microns, though variations in the deep 
telluric absorption features shortward of 2.05 $\mu$m limit the useful 
range to $\lambda$$\ga$2.05 $\mu$m. Wavelength calibration was achieved 
with respect to standard Ne lamps, and telluric correction was achieved by 
observing a bright F star, HD 26784.

All spectra were obtained in an ABBA nod pattern to allow for sky 
subtraction. As for the optical spectra above, the infrared spectra were 
processed using standard IRAF tasks, and we used the IRAF task SPLOT to 
measure equivalent widths of spectral lines. We summarize the observations 
in Table 4.

\subsection{Imaging}

We obtained high-precision astrometric measurements for a subset of our 
candidate companion sample in the course of several adaptive optics 
observing runs at the Keck-2 10m telescope and the Palomar Hale 200'' 
telescope. All observations were obtained using the facility adaptive 
optics imagers, NIRC2 and PHARO. Most of our targets were observed using 
natural guide star adaptive optics (NGSAO), but several faint targets were 
observed at Keck with laser guide star adaptive optics (LGSAO; Wizinowich 
et al. 2006). We also observed a small number of targets with 
seeing-limited imaging during periods of moderate cloud cover that 
prevented the use of adaptive optics. We summarize all of these 
observations in Table 5.

For faint targets, images were obtained using the $K'$ filter 
at Keck or the $K_s$ filter at Palomar. For brighter targets, 
we used the $Br\gamma$ filter, which attenuates flux by a 
factor of $\sim$10 relative to broadband $K$ filters. All of 
our NIRC2 observations were obtained in the 10 mas pix$^{-1}$ 
or 40 mas pix$^{-1}$ modes, depending on whether the binary 
could fit in the narrow-frame FOV (10.18\arcsec) or required 
the wide-frame FOV (40.64\arcsec). All PHARO observations were 
obtained with the 25 mas pix$^{-1}$ mode ($FOV=25.6\arcsec$). 
All Palomar image sets were obtained in a five-point box dither 
pattern. At Keck, all NGSAO observations and early LGSAO 
observations were obtained in a three-point box dither pattern 
(designed to avoid the bottom-left quadrant, which suffers from 
high read noise); later LGSAO observations were obtained in a 
diagonal two-point dither pattern because experience showed 
that dithers degrade the AO correction until several exposures 
have been taken with the Low-Bandwidth Wavefront Sensor, imposing a 
significant overhead.

Most of the targets are relatively bright and require very short 
integration times to avoid nonlinearity, so most exposures were taken in 
correlated double-sampling mode, for which the array read noise is 38 
electrons read$^{-1}$. Where possible, we observed targets in multiple 
correlated double-sampling mode, where multiple reads are taken at the 
beginning and ending of each exposure; this choice reduces the read noise by 
approximately the square root of the number of reads. In most cases, the 
read noise is negligible compared to the signal from the science targets. 
The read noise is negligible ($<$10 electrons read $^{-1}$) in all PHARO 
exposures.

The data were flat-fielded and dark- and bias-subtracted using 
standard IRAF procedures. The NIRC2 images were distortion-corrected 
using new high-order distortion solutions (Cameron 2008) that deliver 
a significant performance increase as compared to the solutions 
presented in the NIRC2 pre-ship 
manual\footnote{http://www2.keck.hawaii.edu/realpublic/inst/nirc2/}; 
the typical absolute residuals are $\sim$4 mas in wide camera mode and 
$\sim$0.6 mas in narrow camera mode. The PHARO images were 
distortion-corrected using the solution derived by Metchev (2005). We 
adopted the NIRC2 narrow-field plate scale (9.963 $\pm$0.003 mas 
pix$^{-1}$) and y-axis PA (in degrees east of north; +0.13 $\pm$0.01 
deg) reported by Ghez et al. (2008). As we will report in a future 
publication (Kraus, Ireland, et al., in prep), we then used 
observations of the M5 core (e.g. Cameron et al. 2009) to extrapolate 
corresponding values for the NIRC2 wide-field camera (39.83 $\pm$0.04 
mas pix$^{-1}$ and +0.34$\pm$0.02 deg) and the PHARO narrow-field 
camera (25.19$\pm$0.04 mas pix$^{-1}$ and +2.15$\pm$0.10 deg, assuming 
the Cassegrain ring is set at +335 deg). The rotation for PHARO might 
change over time and this value has only been confirmed for 2007, so 
new calibrations will be needed for any other epochs. The values for 
PHARO also differ from those adopted in Kraus et al. (2008), where we 
used old values of the plate scale and rotation, so we have 
recalibrated the previous results to match the updated values.

We measured photometry and astrometry for our sources using the 
IRAF package DAOPHOT (Stetson 1987). For systems with small or 
moderate separations, we used the PSF-fitting ALLSTAR routine. 
For systems with wider separations, where anisoplanatism 
produced significantly different PSFs, we used the PHOT 
package. We analyzed each frame separately in order to estimate 
the uncertainty in individual measurements and to allow for the 
potential rejection of frames with inferior AO correction; our 
final results represent the mean value for all observations in 
a filter. For observations where the primary star was single or 
the secondary was close to on-axis ($\rho$$\la$5\arcsec), we 
used that source to produce individual template PSFs for each 
image. In the few cases where a source was itself a close 
binary, we measured photometry and astrometry for each close 
component using the PSF reconstruction technique that we 
described in Kraus \& Hillenbrand (2007b), then combined the values to 
find the photocenter.

We calibrated our photometry using the known 2MASS $K_s$ magnitudes for 
each of our science targets; in cases where the binary system was not 
resolved in the 2MASS PSC, we invoked the estimated $K_s$ magnitudes for 
each component from our discovery survey (KH07a). Our broadband photometry 
was obtained using both $K'$ and $K_s$ filters, but previous comparisons 
have shown that the filter zero points differ by $\la$0.01 mag for objects 
with typical stellar colors (Carpenter 2001; Kim et al. 2005). We tested 
the systematic uncertainty for late-type objects by convolving template 
spectra from the IRTF Spectral Library (Rayner et al., in prep) with the 
filter profiles; our results show that the zero point for the $K'$ filter 
is $\sim$0.05 mag fainter than for $K_s$ at a spectral type of M7. The 
midpoint of the narrow $Br\gamma$ filter is very close to the midpoint of 
typical $K$ filters (2.166$\mu$), so its calibration uncertainty should be 
similar.

The calibration process could introduce larger systematic uncertainties 
($\sim$0.1--0.2 mag) if any of the sources are variable, as many pre-main 
sequence stars tend to be, but these cases can be identified if the 
calibrated flux ratios for candidate binary components do not agree with 
previous measurements. For systems observed with the $Br\gamma$ filter, 
there could also be a systematic error if one component shows line 
emission (likely due to accretion) while the other does not; the magnitude 
of the error would then depend on the line flux relative to the continuum 
flux.

Finally, we note that one target (the candidate companion to USco80) was 
resolved to be a close equal-flux pair. Our analysis for the system 
reflects this discovery, and we will describe this observation in more 
detail in a future publication that summarizes our ongoing survey of the 
multiplicity of very low-mass stars and brown dwarfs.

\begin{deluxetable}{lccrrl}
\tabletypesize{\tiny}
\tablewidth{0pt}
\tablecaption{Imaging Observations}
\tablehead{\colhead{Candidate} & \colhead{Telescope/} & 
\colhead{$T_{int}$} & \colhead{Scale} & 
\colhead{Epoch(JD-}
\\
\colhead{Companion} & \colhead{Mode} & \colhead{(sec)} &
\colhead{(mas)} & \colhead{2400000)}
}
\startdata
2M04080771+2807373&Keck/NGS&40&40&54069\\
GSC 06213-01459 B&Keck/NGS&120&10&54187\\
2M16101888-2502325&Keck/NGS&160&10&54188\\
2M15410726-2656254&Pal/NGS&50&25&54198\\
2M15554839-2512174&Pal/NGS&297&25&54198\\
2M16151239-2420091&Keck/LGS&20&40&54188\\
2M16182365-2338268&Pal/Seeing&1427&25&54199\\
2M16213638-2355283&Pal/Seeing&1308&25&54199\\
2M15583621-2348018&Keck/LGS&120&40&54188\\
\enddata
\end{deluxetable}

\subsection{Archival Astrometry}

We retrieved relative astrometry for our wide companion sample 
from several all-sky imaging surveys: the Two-Micron All-Sky 
Survey (2MASS; Skrutskie et al. 2006), the Deep Near Infrared 
Survey (DENIS; Epchtein et al. 1999), and the United States 
Naval Observatory B1.0 survey (USNOB; Monet et al. 2003). The 
DENIS and 2MASS source catalogues are based on wide-field 
imaging surveys conducted in the optical/NIR ($IJK$ and $JHK$, 
respectively) using infrared array detectors, while USNOB is 
based on a digitization of photographic plates from the Palomar 
Observatory Sky Surveys.

In our discovery survey (KH07a), we presented 2MASS astrometry for each 
filter that was measured directly from the processed atlas images, so we 
have adopted those values. We extracted DENIS astrometry from the source 
catalog, which contains the average value for all three filters. The USNOB 
source catalog reports processed astrometry as well as individual 
astrometric measurements for each epoch; we have chosen to work with the 
individual measurements since it is unclear how the USNOB astrometric 
pipeline weighted individual measurements or rejected potentially 
erroneous measurements.

Both 2MASS and DENIS quote astrometric uncertainties of 70-100 mas for 
individual sources spanning the brightness range of our sample, while 
USNOB reports uncertainties of $\sim$200-300 mas in each epoch. However, 
the quoted uncertainties include significant systematic terms resulting 
from the transformation to an all-sky reference frame. We have conducted 
tests with our known binary systems with existing high-precision 
measurements (Table 3) which suggest that narrow-angle astrometry on 
angular scales of $<$1\arcmin\, is accurate to $\sim$70 mas for 
2MASS/DENIS and 100-200 mas for USNOB, depending on brightness, so we 
adopt these lower values as the astrometric uncertainties for all 
measurements.

We also collated all of the astrometric observations reported in the 
literature for our wide companion sample. Most of these measurements were 
obtained using high-resolution imaging techniques: lunar occultation 
interferometry, speckle interferometry, and adaptive optics imaging. 
However, some were also obtained with seeing-limited imaging. In each 
case, we adopt the uncertainties reported in the literature, but it is 
unclear in many cases whether all possible sources of systematic error 
(such as geometric distortion or unresolved multiplicity) have been 
assessed.

\subsection{Stellar and Companion Properties}

Stellar properties can be difficult to estimate from observed properties, 
particularly for young stars, since pre-main-sequence stellar evolutionary 
models are not well-calibrated. The masses of a given sample could be 
systematically uncertain by as much as 20\% (e.g. Hillenbrand \& White 
2004), and individual masses could be uncertain by factors of 50\% or more 
due to unresolved multiplicity or the intrinsic variability that accreting 
young stars often display. These caveats suggest that any prescription for 
determining stellar properties should be treated with caution.

We estimated the properties of our sample members using the methods 
described in our original discovery survey (KH07a). This procedure 
calculates component masses by combining the 2- or 5-Myr isochrones of 
Baraffe et al. (1998) and the M dwarf temperature scale of Luhman et al. 
(2003) to convert observed spectral types to masses. Relative properties 
(mass ratios $q$) are calculated by combining the Baraffe isochrones and 
Luhman temperature scale with the empirical NIR colors of Bessell \& Brett 
(1998) and the K-band bolometric corrections of Leggett et al. (1998) to 
estimate $q$ from the observed flux ratio $\Delta$$K$. The observed flux 
ratio is not sensitive to the distance or extinction for a system (unless 
differential extinction is present), so the relative system properties 
should not be affected by these potential sources of error. We also used 
these techniques to estimate masses for all single stars and confirmed 
binary pairs in our sample.

For all binary systems without spatially resolved spectra, we have adopted 
the previously-measured (unresolved) spectral type for the brightest 
component and inferred its properties from that spectral type. This 
assumption should be robust since equal-flux binary components will have 
similar spectral types and significantly fainter components would not have 
contributed significant flux to the original discovery spectrum. The 
properties of all fainter binary components were then inferred using the 
methods described in the previous paragraph. When we compute 
mass-dependent properties (mass ratios and total system masses) for our 
samples, we sum the masses of all sub-components of our wide "primary" and 
"secondary".

Projected spatial separations are calculated assuming the mean distance 
for each association, $\sim$145 pc (de Zeeuw et al. 1999; Torres et al. 
2007). If the total radial depth of each association is equal to its 
angular extent ($\sim$15$^o$ or $\sim$40 pc), then the unknown depth of 
each system within its association implies an uncertainty in the projected 
spatial separation of $\pm$15\%. The systematic uncertainty due to the 
uncertainty in the mean distance of each association is negligible in 
comparison ($\la$5\%).

\section{Results}

\subsection{Optical Spectroscopy}

The spectra show that our candidate companions can be divided into three 
groups: background dwarfs, background GK giants, and young association 
members. We plot the corresponding spectra in Figures 1-3, respectively, 
and we summarize our spectral classifications in Table 8.

\subsubsection{Background Dwarfs}

 \begin{figure}
 \plotone{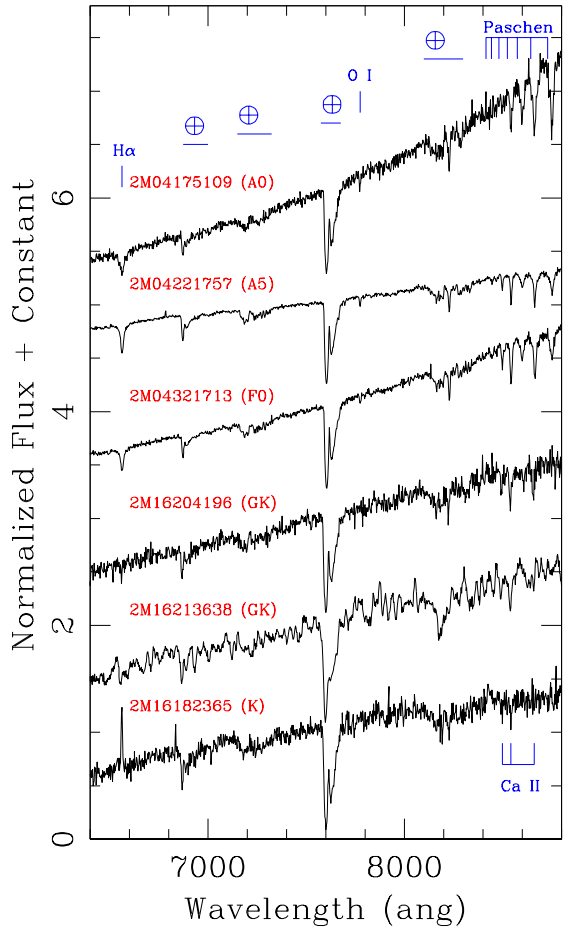}
 \caption{Six field dwarfs that are located behind Taurus or Upper Sco. 
The top three spectra show clear absorption from the Paschen-14 and -12 
lines, indicating that the sources are background A-F stars. The next two 
spectra show absorption from the Ca II infrared triplet, but no absorption 
features from the Paschen series or from TiO bands, indicating that the 
sources are G-K stars. Finally, the bottom star shows $H\alpha$ emission 
that might indicate accretion (and youth), but it could also indicate the 
presence of an active M dwarf companion. In all cases, the stars are too 
faint for their spectral type to be members, indicating that they are 
located behind the associations. Finally, we note that the spectrum for 
2M16213638-2355283 was smoothed with a five-pixel average to emphasize the 
absence of broad TiO absorption bands, so most apparently narrower 
features (i.e. the apparent absorption feature at 8200A) are noise 
artifacts. All relevant spectral features and atmospheric absorption 
bands have been labeled.}
 \end{figure}

Early-type A and F dwarfs are easily identified by the presence of the 
Paschen series at $\ga$8400 angstroms, and specifically by the Paschen-12 
and -14 lines at 8595 and 8748 angstroms. The Paschen sequence fades and 
the CaII infrared triplet grows between late A and late F, so the relative 
depths of Paschen-14 and the CaII triplet provide an excellent diagnostic 
for temperature in this range. We identified three sources with these key 
features, and we determined approximate spectral types for each source by 
comparing our spectra to the standard stars of Torres-Dodgen \& Weaver (1993) and 
Allen \& Strom (1995).

All three of the A-F stars that we observed are faint ($K\sim$11-12) and 
reddened to varying degrees ($J-K\sim0.8$ for 2M04321713+2421556 and 
$J-K\sim1.4$ for the other two stars). Assuming their dereddened colors 
are $J-K\sim0.0$, these colors suggest extinctions of $A_K\sim0.5$ and 
$A_K\sim1.0$, respectively, according the reddening law of Schlegel et 
al. (1997). The corresponding dereddened apparent magnitudes are far too 
faint ($K\ga10$) to denote association members, suggesting that these 
stars are located beyond the association at a distance of $\sim$1 kpc.

We also identified two additional candidates, 2M16204196-2426149 and 
2M16213638-2355238, that also appear to be reddened dwarfs. The Ca II 
infrared triplet is clearly detected for the former, but there is no 
convincing evidence of the Paschen series or TiO absorption bands, 
suggesting that it has a spectral type between early G and mid K. As we 
will describe in the next subsection, background giants possess a 
significant CN band at 7900A that this star appears to lack, suggesting 
that it is a dwarf. The spectral type of 2M16213638-2355238 is more 
difficult to assess due to the higher noise, but the absence of the TiO 
absorption bands suggests a spectral type of $<$M0.

Like the A-F stars, these candidate companions are faint and reddened 
($K=11.5$ to 12.5, $J-K\sim1.4$, $H-K\sim0.35$). If they have the 
dereddened colors of a G-K star ($J-K\sim0.5$, $H-K\sim0.1$; Bessell \& 
Brett 1988), then these colors suggest an extinction of $A_K\sim$0.6 and 
corresponding dereddened apparent magnitudes of $K\sim$11 to 12. This 
flux is far too faint to identify either source as a G-K type Upper Sco 
member, but is approximately consistent with a dwarf at a distance of 
$\sim$200-500 pc. This interpretation would normally be suspect for an 
object located behind Upper Sco since most of the interstellar material 
in the region has been dispersed, but both of these objects are located 
close to the edge of Ophiuchus, so the presence of interstellar material 
is not surprising. For example, Bouy et al. (2007) noted that extinction 
is locally higher along the line of sight to DENIS162041.5-242549.0 
($A_V=3.3$ or $A_K\sim0.3$).

Finally, 2M16182365-2338268 appears to be a K dwarf in the background of 
the association; the absence of TiO absorption at 6700 angstroms and the 
CaII infrared triplet at 8500 angstroms suggest that the spectral type 
is not $\ga$K7 or $\la$K0, and the shape of the continuum indicates 
moderate reddening that would not occur if it were in the foreground. As 
in the previous cases, it is faint and red ($K=12.25$, $J-K=1.31$, 
$H-K=0.34$). If its intrinsic colors are $J-K=0.6$ and $H-K=0.12$, then 
the apparent colors suggest an extinction of $A_V\sim$4 and a dereddened 
apparent magnitude of $K\sim11.8$. This flux places the candidate well 
below the association sequence, but is consistent with a K5V star at a 
distance of $\sim$300 pc.

The presence of moderate $H\alpha$ emission makes this 
identification somewhat arguable since $H\alpha$ emission is a 
key indicator of accretion (and youth). However, it could also 
indicate the presence of an (unresolved) active M dwarf 
companion, so it is not conclusive by itself. As I will 
describe in Section 4.2, this candidate's relative proper 
motion is also inconsistent with comovement, which supports the 
spectroscopic identification of this candidate as a nonmember.

\subsubsection{Background Giants}

 \begin{figure}
 \plotone{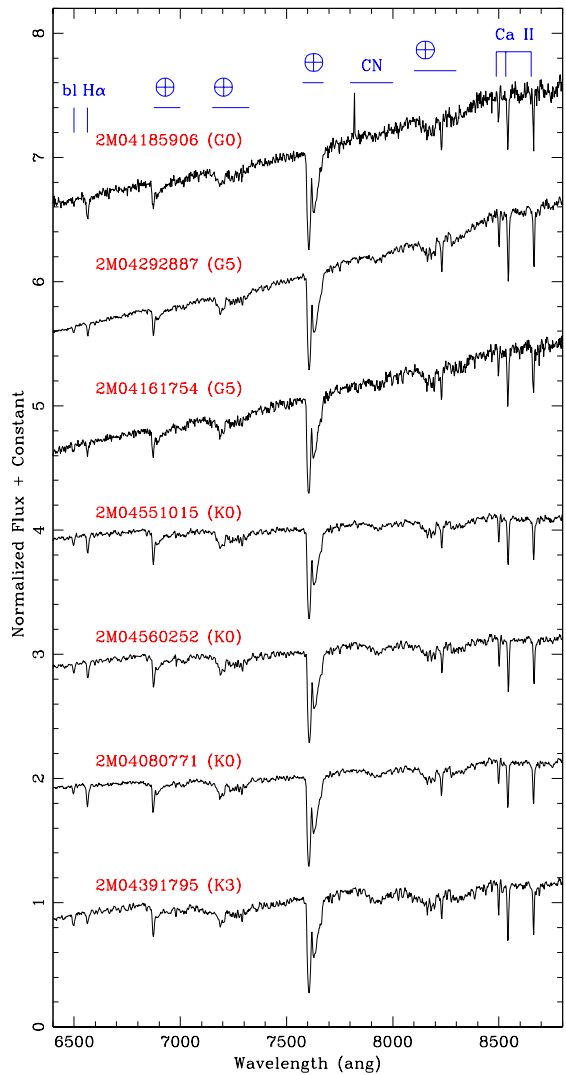}
 \caption{Eight field giants that are located behind Taurus or Upper Sco. 
All spectra show absorption from the CN band at 7900 angstroms and deep, 
narrow absorption lines in the CaII infrared triplet, indicating that the 
sources are giants. Given their brightness, all are located behind the 
associations, consistent with the significant reddening seen for several 
of them. The approximate spectral type has been estimated based on the 
ratio of line strengths for H$\alpha$ and the blend of several metal lines 
at 6497 $\AA$ (denoted bl). All relevant spectral features and 
atmospheric absorption bands have been labeled.}
 \end{figure}

Background giants can also be easily identified, most readily 
by the presence of a broad CN absorption band at 7900 
angstroms. It has long been known (e.g. White \& Wing 1978; 
MacConnell et al. 1992; Torres-Dodgen \& Weaver 1993) that this CN band 
is extremely sensitive to luminosity class: very deep for 
supergiants, shallow for giants, and completely absent for 
dwarfs. This result suggests that any source with detectable CN 
absorption is a luminous, distant background giant rather than 
an association member. The depth of the CN band has been 
characterized via the narrowband photometric system first 
described by Wing (1971), but that system is calibrated using 
fluxes beyond the red limit of our spectra, so we could not 
implement it without significant modification. Our only goal is 
to identify background giants and remove them from further 
consideration, so we opted simply to identify the presence of 
CN absorption by visual inspection. The deep, narrow absorption 
lines in the CaII infrared triplet also support our 
identifications.

There are few spectral type indicators in this wavelength range 
for G-K stars, and most are poorly calibrated, but we have used 
them to assess approximate spectral types with respect to the 
standard stars of Torres-Dodgen \& Weaver (1993) and Allen \& Strom 
(1995). We can rule out spectral types of $\ge$K4 for all of 
these stars since TiO absorption appears and grows with 
decreasing temperature. The relative depths of H$\alpha$ and 
the metal blend at 6497 angstroms gradually change across the G 
and K spectral types, with the blend appearing at $\sim$G0 and 
equaling the depth of H$\alpha$ at K3, so we used their 
relative depths to assess stars as spectral type G0, G5, K0, or 
K3. Residual absorption in the Paschen-14 line can also persist 
as late as $\sim$G5, which also helped us to distinguish 
between G giants and K giants.

\subsubsection{Young Stars}

 \begin{figure*}
 \epsscale{1.00}
 \plotone{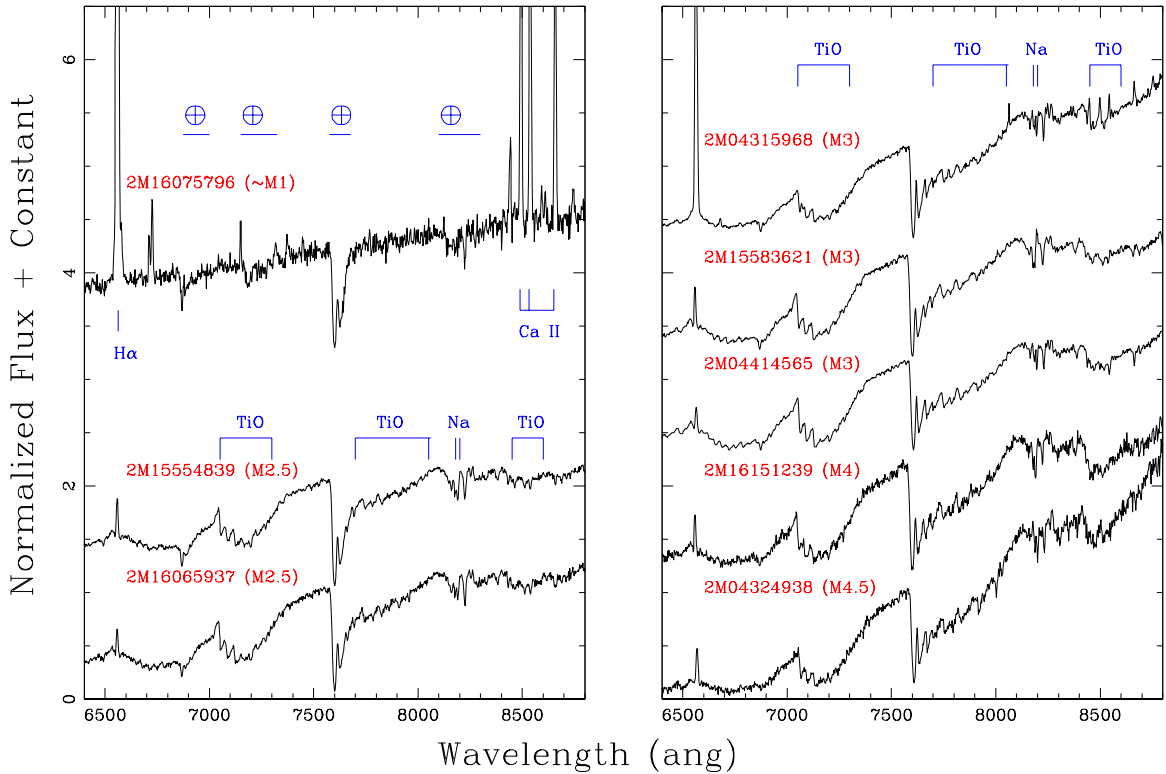}
 \caption{Spectra for eight new association members. One star in Upper Sco 
(2M16075796; top left) shows extremely strong emission at H$\alpha$ and Ca 
II, consistent with strong accretion. The other seven stars are M dwarfs 
with low surface gravity (as measured from the Na-8189 doublet), which 
indicates that these stars have not yet contracted to the zero-age main 
sequence. All strong spectral features and atmospheric absorption bands 
have been labeled. We find that 2M16075796 also has numerous emission 
lines which are usually associated with accretion-driven jets: [N II] 
6584, [S II] 6717/6731, [Fe II] 7155, [Ca II] 7323, [Ni II] 7378, OI 8446, 
and the Paschen series. Emission from the CaII infrared triplet indicates 
that 2M04315968 is accreting as well.}
 \end{figure*}

Stellar youth is most commonly inferred from three major classes of 
spectroscopic features: accretion signatures like H$\alpha$, HeI, and CaII 
emission, low-gravity diagnostics like shallow absorption from the Na-8189 
doublet, or lithium absorption at 6708A. The spectral resolution of our 
observations ($R\sim$1200) can detect lithium only at very high S/N. 
Surface gravity can be assessed for stars later than M1 by the depth of 
the Na-8189 doublet, but all of the standard gravity indicators for K 
stars have wavelengths shorter than the blue limit of our spectra, so for 
K stars, our only option is to search for accretion signatures. We 
identified one K-M star based on its accretion and 7 M stars based on 
their surface gravity.

The optical classification of M stars is very straightforward due to 
their numerous and distinct molecular bands. Across the wavelength range 
of our spectra, early M stars are most distinctly classified by the 
depth of the TiO bandhead at 7050 angstroms, while mid-M stars are more 
distinctively classified by the depth of the TiO bandhead at 8500 
angstroms. We have assessed all spectral types using the spectral 
indices TiO$_{7140}$ and TiO$_{8465}$ (Slesnick et al. 2006a), supported 
by a visual inspection of each spectrum. We adopted our spectral 
standards from a list originally observed by Slesnick et al. 
(2006a,2006b) using DBSP with identical instrument settings. We assessed 
the surface gravity using the Na$_{8189}$ index developed by Slesnick et 
al. (2006a), confirming that each source was young by comparing its 
TiO$_{7140}$ and Na$_{8189}$ indices to the dwarf, young star, and giant 
results that they reported for their survey. As we show in Figure 3, 
qualitative inspection of the Na-8189 doublet for all seven M stars in 
our sample indicated that it was shallower than the field, but roughly 
similar to known members of Taurus or Upper Sco.

The other young star in our optical spectroscopy sample, 
2M16075796-2040087, is easily identified by the obvious presence of 
accretion signatures; as we demonstrate in Figure 3, it shows tremendous 
H$\alpha$ emission ($EW=-357$ angstroms) and significant emission from the 
CaII infrared triplet (-30.0, -31.8, and -25.4 angstroms at 8500, 8542, 
and 8664 angstroms). Several other emission line features indicate that a 
jet is being driven by the accretion process. The absence of absorption 
features makes it impossible to place an early limit on the star's 
spectral type. Its $J$ band magnitude ($J=11.06$), which should be least 
affected by optical veiling or NIR disk emission, is roughly consistent 
with other M0-M2 members, so we have assigned a preliminary spectral type 
of M1. Emission from the CaII infrared triplet indicates that 
2M04315968+1821305 is accreting as well, but it lacks the forbidden 
emission lines that are present for 2M16075796-2040087.

\subsection{Near-Infrared Spectroscopy}

 \begin{figure*}
 \epsscale{1.00}
 \plotone{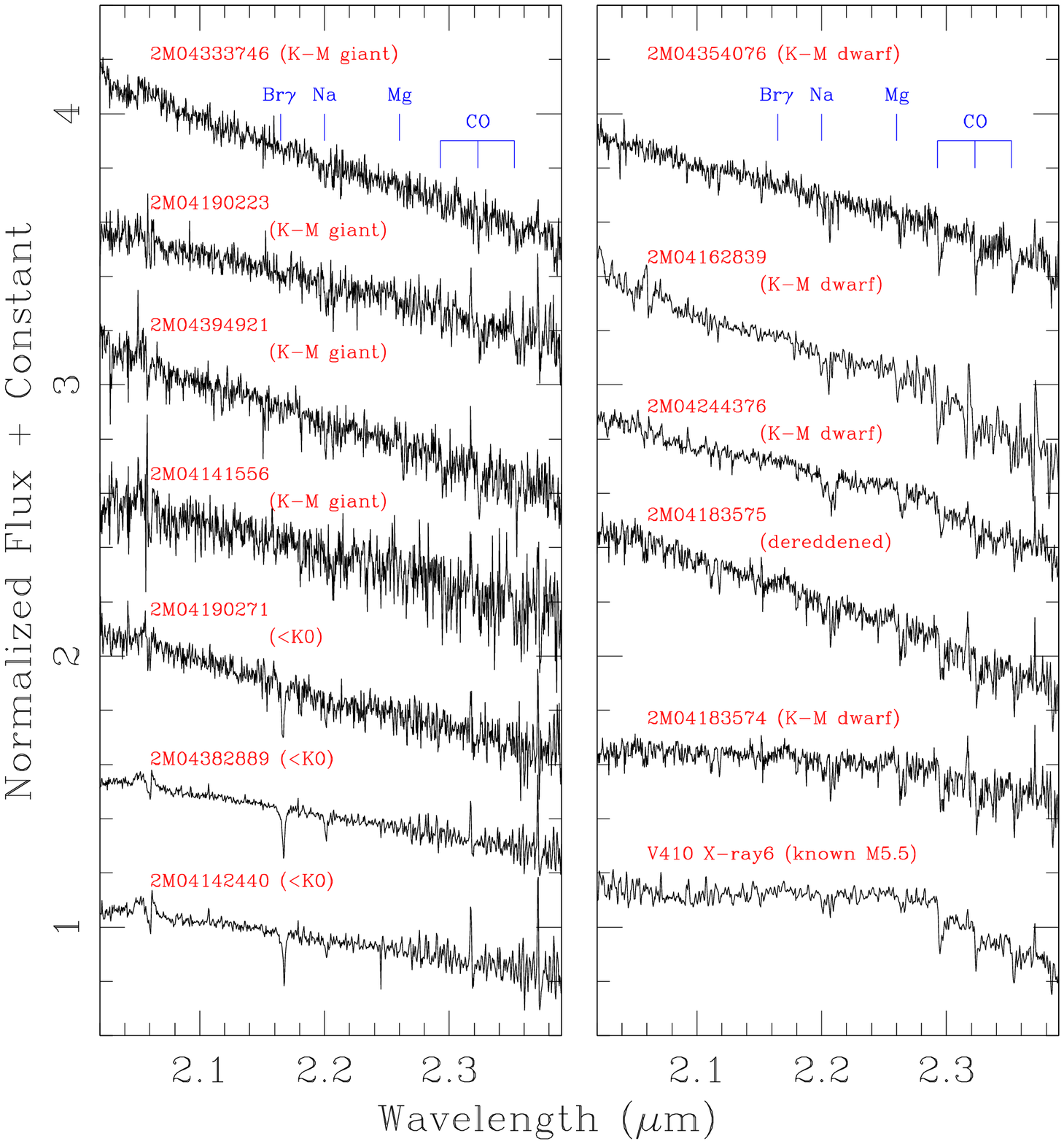}
 \caption{K-band spectra for 11 candidate companions and one known 
Taurus member (V410 X-ray6; M5.5). The three companions in the bottom 
left all possess significant Br$\gamma$ absorption, which indicates 
that the sources are background early-type stars. The rest of the 
candidates appear to be field K-M stars, divided between dwarfs and 
giants. All relevant spectral features have been labeled.}
 \end{figure*}

K-band spectra include several key features that are useful for stellar 
classification (e.g. Slesnick et al. 2004). The Br$\gamma$ absorption line 
at 2.166 microns is ubiquitous for all stars with spectral types earlier 
than K, but disappears entirely by mid-K. Conversely, the CO absorption 
bandheads at $\ge$2.3 microns are present (with similar depths) for all 
late-type stars, but they start to weaken at mid-K and disappear entirely 
for stars earlier than late-G. Both Br$\gamma$ and the CO bandheads can 
also appear in emission for young stars. A broad steam absorption band at 
$\la$2 microns is also a key indicator for identifying M stars with low 
S/N spectra since it grows with decreasing temperature, though its depth 
is gravity sensitive; at a given spectral type, it is deeper for dwarfs 
than for giants. The depths of the Na and Mg doublets (at 2.2 and 2.26 
microns) are also useful for distinguishing the luminosity classes of 
stars because they increase with surface gravity, though the 
identification requires good $S/N$. Finally, our efforts are aided 
significantly by the faintness of our targets; any candidates with 
spectral types earlier than mid-M must fall significantly below the 
association sequence on an HR diagram.

We plot all of our K-band spectra in Figure 4, including a spectrum of 
the known member V410 X-ray 6 (M5.5) to demonstrate the expected 
morphology for young low-mass stars or brown dwarfs. Three of our 
candidate companions show clear Br$\gamma$ absorption, indicating that 
the sources are background stars with early spectral types ($<$K0). The 
other 8 targets all show some degree of absorption in the CO bandheads, 
indicating spectral types of K-M. However, seven of these targets 
clearly show no evidence of steam absorption, indicating that the 
sources are either background K-M giants or dwarfs with spectral types 
$\la$M1. In either case, all sources are too faint for their dereddened 
magnitudes to fall along the Taurus sequence, so we have divided them 
into giants or dwarfs based on the strength of their Na and Mg doublets. 
These classifications are preliminary due to the low S/N of many 
spectra, but they are sufficient to rule out the possibility of 
membership.

The eighth K-M star (2M04183574+2830254, the neighbor of V410 X-ray 2) 
is significantly reddened, which complicates its classification. Its NIR 
colors ($J-K=4$, $H-K=1.5$) suggest a visual extinction of $A_V\sim$20 
(matching the value for V410 X-ray 2 itself, based on its 2MASS colors), 
so we removed this effect with the IRAF task deredden. As we show in 
Figure 4, the dereddened spectrum possesses significant Na and Mg 
absorption, but no steam absorption, suggesting that it is a field dwarf 
with spectral type $\la$M1 and that it is located behind the material 
that obscures V410 X-ray 2.

\subsection{Astrometry}

The other standard method for confirming candidate binary 
companions is to test for common proper motion. This test is 
less useful for young stars because other (gravitationally 
unbound) association members are also comoving to within the 
limits of our observational uncertainties. However, proper 
motion analysis can still be used to eliminate foreground and 
background stars that coincidentally fall along the association 
color-magnitude sequence but possess distinct kinematics.

In Table 6, we list the relative astrometric measurements for each 
candidate binary pair that we obtained from the literature and from our 
observations. We computed relative proper motions by using a weighted least 
squares fit to determine the relative motion in each dimension, rejecting 
the worst-fitting measurement if it differed from the fit by more than 
3$\sigma$ (where $\sigma$ is the observational error, not the dispersion in 
the fit). A cutoff of 3$\sigma$ in a bivariate normal distribution 
corresponds to a confidence level of $\sim$99\%, so we do not expect many 
valid measurements to be flagged. We did not reject multiple measurements 
that differ by $>$3$\sigma$ because the high scatter could indicate an 
astrophysical source for the poor astrometric fit (such as further 
unresolved multiplicity).

In Table 7, we list the proper motions that we derived for each candidate 
companion. In Figure 5, we plot the relative proper motion of each 
candidate companion with respect to its corresponding known association 
member. For each association, there are two major concentrations: one 
group centered on the origin, corresponding to comoving young association 
members, and one group centered on the inverse proper motion for that 
association, corresponding to nonmoving background stars. There are also 
several objects which fall outside both concentrations, which could 
correspond to either independently moving field dwarfs or objects with 
erroneous astrometry. We also specifically mark those objects which were 
spectroscopically confirmed to be members or nonmembers; all 5 
spectroscopic members and only 1 of 9 confirmed nonmembers fall in the 
cluster of sources centered on the origin ($\Delta$$\mu$$\la$12 mas 
yr$^{-1}$).

We find that 7 of the 15 candidates without spectra fall inside this 
limit, which suggests that no more than $\sim$1 of them is also comoving 
by chance. We therefore treat all candidates which are comoving to $\la$12 
mas yr$^{-1}$ as likely companions and all other candidates as likely 
contaminants. We have opted not to use more rigorous selection criteria 
(based on our formal uncertainties) because the distribution of likely 
members seems too large for our uncertainties to be accurate, even among 
our spectroscopically confirmed subsample alone. Given the many 
astrophysical and observational sources of systematic uncertainty that can 
influence high-precision astrometry, all of our proper motion 
uncertainties are probably underestimated by a factor of $\sim$2 (the 
multiplier needed to bring our uncertainties in line with the observed 
scatter).

We list all of our membership assessments in Table 8, denoting likely 
companions and likely contaminants with "Y?" and "N?", respectively. 
Spectroscopic membership analysis should generally supercede these 
determinations, and given the value of directly determining a companion's 
stellar properties, followup observations for all of these likely 
companions should be a high priority. However, the existing data should 
suffice for studying the bulk properties of our sample.

\begin{deluxetable*}{lllccl}
\tabletypesize{\scriptsize}
\tablewidth{0pt}
\tablecaption{Astrometric Data}
\tablehead{\colhead{Known Member} & \colhead{Candidate Companion} & 
\colhead{Epoch} & \colhead{Sep} & \colhead{PA} & \colhead{Ref}
\\
\colhead{} & \colhead{} & \colhead{(JD-2400000)} & \colhead{(mas)} & \colhead{(deg)}
}
\startdata
\multicolumn{6}{c}{New}\\
2M04080782+2807280&2M04080771+2807373&54069&9508$\pm$15&351.15$\pm$0.02&Keck-NGS\\
DG Tau&2M04270370+2606067&54434&16322$\pm$29&235.35$\pm$0.11&Palomar-NGS\\
GSC 06213-01459&GSC 06213-01459 B&54187&3213$\pm$2&306.3$\pm$0.02&Keck-NGS\\
GSC 06784-00997&2M16101888-2502325&54188&4896$\pm$2&241.24$\pm$0.02&Keck-NGS\\
GSC 06785-00476&2M15410726-2656254&54198&6270$\pm$10&82.65$\pm$0.1&Palomar-NGS\\
RXJ 1555.8-2512&2M15554839-2512174&54198&8877$\pm$14&319.73$\pm$0.1&Palomar-NGS\\
RXJ 1555.8-2512&2M15554788-2512172&54198&14524$\pm$23&299.27$\pm$0.1&Palomar-NGS\\
SCH161511.15-242015.56&2M16151239-2420091&54188&17885$\pm$22&70.24$\pm$0.07&Keck-LGS\\
SCH161825.01-233810.68&2M16182365-2338268&54199&24510$\pm$50&229.87$\pm$0.12&Palomar-Seeing\\
USco80&2M15583621-2348018&54188&12274$\pm$23&15.59$\pm$0.04&Keck-LGS\\
\multicolumn{6}{c}{Archival}\\
2M04080782+2807280&2M04080771+2807373&50781&9432$\pm$70&351.0$\pm$0.4&2MASS H\\
2M04080782+2807280&2M04080771+2807373&50781&9420$\pm$70&350.7$\pm$0.4&2MASS J\\
2M04080782+2807280&2M04080771+2807373&50781&9416$\pm$70&351.7$\pm$0.4&2MASS K\\
2M04080782+2807280&2M04080771+2807373&35403&7850$\pm$200&353.2$\pm$1.5&USNOB B1\\
2M04080782+2807280&2M04080771+2807373&48896&8620$\pm$200&351.0$\pm$1.3&USNOB B2\\
2M04161885+2752155&2M04161754+2751534&50782&28063$\pm$70&218.3$\pm$0.1&2MASS H\\
2M04161885+2752155&2M04161754+2751534&50782&28033$\pm$70&218.3$\pm$0.1&2MASS J\\
2M04161885+2752155&2M04161754+2751534&48896&27760$\pm$200&217.5$\pm$0.4&USNOB B2\\
2M04161885+2752155&2M04161754+2751534&50337&27970$\pm$200&218.7$\pm$0.4&USNOB I2\\
2M04161885+2752155&2M04161754+2751534&35403&28630$\pm$200&215.9$\pm$0.4&USNOB R1\\
2M04161885+2752155&2M04161754+2751534&47827&28000$\pm$200&218.3$\pm$0.4&USNOB R2\\
\enddata
\tablecomments{The full version of this table will be available in the online 
version of ApJ.}
\end{deluxetable*}

\begin{deluxetable*}{llrrr}
\tabletypesize{\tiny}
\tablewidth{0pt}
\tablecaption{Companion Kinematics}
\tablehead{\colhead{Known Member} & \colhead{Candidate Companion} & 
\multicolumn{2}{c}{Relative Motion} & \colhead{$\sigma$$_{\mu}$} \\
\colhead{} & \colhead{} & \colhead{$\mu_{\alpha}$} & \colhead{$\mu_{\delta}$}
& \colhead{(mas yr$^{-1}$)}
}
\startdata
2M04080782+2807280&2M04080771+2807373&-7&24&3\\
2M04161885+2752155&2M04161754+2751534&-15&27&5\\
2M04213460+2701388&2M04213331+2701375&6&17&4\\
CFHT-Tau-21&2M04221757+2654364&-11&5&3\\
CFHT-Tau-7&JH90&-4&20&4\\
DG Tau&2M04270370+2606067&2&22&5\\
FO Tau&2M04382889+2611178&-44&98&5\\
FW Tau&2M04292887+2616483&-2&28&3\\
GM Aur&2M04551015+3021333&-2&25&3\\
HBC 427&2M04560252+3020503&58&-83&5\\
I04385+2550&2M04413842+2556448&8&23&4\\
IS Tau&2M04333467+2609447&-10&33&4\\
V710 Tau AB&2M04315968+1821305&-1&8&3\\
GSC 06213-01459&GSC 06213-01459 B&0&8&5\\
GSC 06784-00997&2M16101888-2502325&-4&10&5\\
GSC 06785-00476&2M15410726-2656254&0&-12&4\\
RXJ 1555.8-2512&2M15554788-2512172&16&19&5\\
RXJ 1555.8-2512&2M15554839-2512174&6&11&5\\
RXJ 1558.8-2512&2M15585415-2512407&10&19&3\\
RXJ 1602.8-2401B&2M16025116-2401502&11&3&5\\
SCH160758.50-203948.90&2M16075693-2039424&19&21&4\\
SCH161511.15-242015.56&2M16151239-2420091&-4&-2&4\\
SCH161825.01-233810.68&2M16182365-2338268&-6&43&5\\
ScoPMS042b&2M16102177-1904021&2.7&0.5&1\\
ScoPMS048&ScoPMS 048 B&-2&3.9&0.8\\
UScoJ160245.4-193037&2M16024735-1930294&55&-2&4\\
UScoJ160700.1-203309&2M16065937-2033047&-5&3&3\\
UScoJ160936.5-184800&2M16093658-1847409&29&13&3\\
UScoJ161031.9-191305&2M16103232-1913085&-9&-5&5\\
USco80&2M15583621-2348018&1&0&3\\
\enddata
\tablecomments{As we discuss in Section 4.3, many of the proper motions 
that rely on high-precision astrometry could be more uncertain due to 
uncorrected systematic effects (such as detector distortion) and 
astrophysical jitter (such as from unresolved high-order multiplicity). A 
factor of $\sim$2 increase in the proper motion uncertainty would bring 
our uncertainties in line with the observed scatter.}
\end{deluxetable*}

 \begin{figure*}
 \epsscale{0.95}
 \plotone{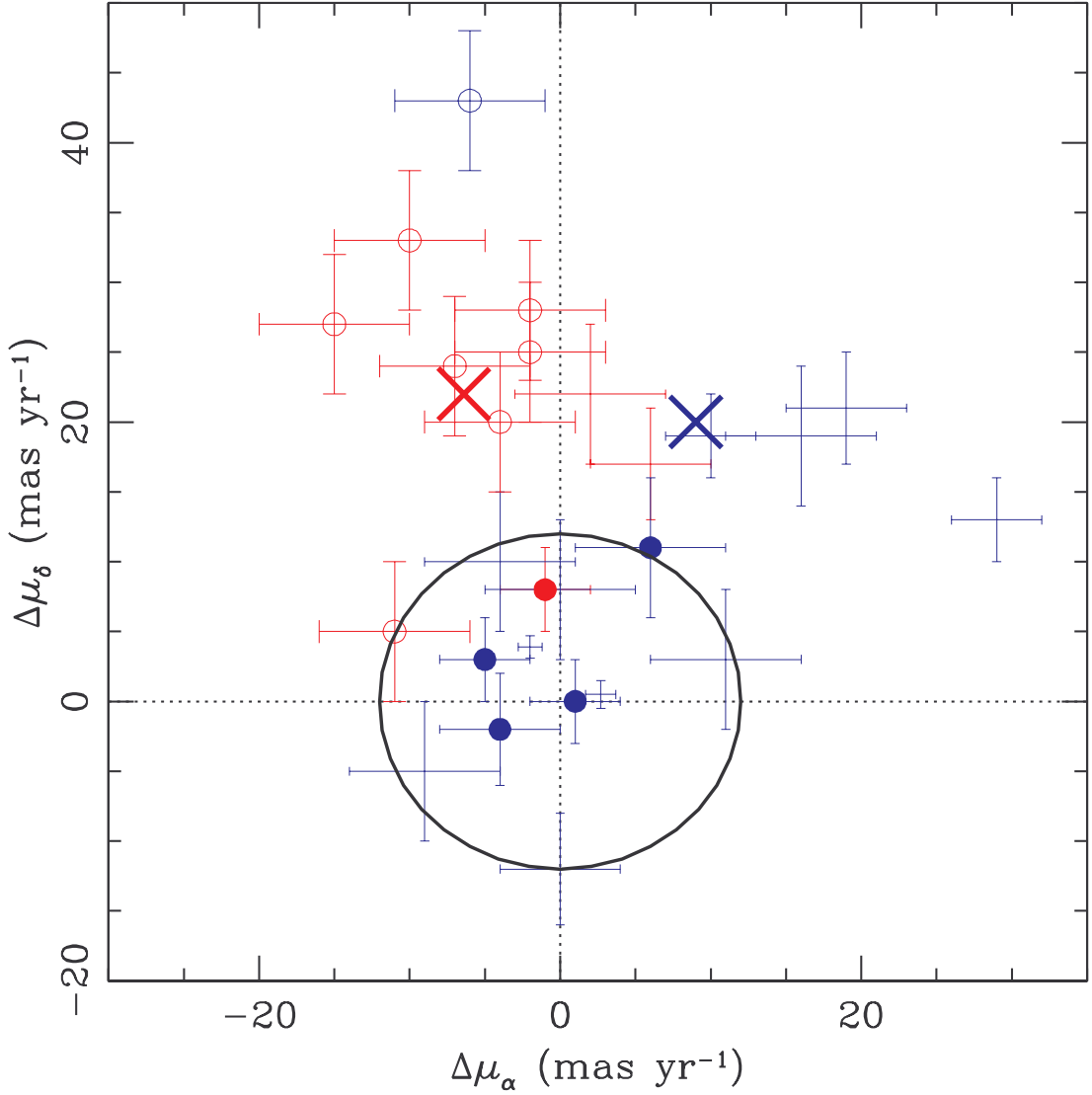}
 \caption{Relative proper motions of each candidate companion with respect 
to the known association member in Taurus (red) and Upper Sco (blue). The 
crosses shown the expected relative motion (in each association) for a 
wide neighbor which is actually a nonmoving background stars; each set of 
association members shows a concentration around this reflex motion 
(denoting nonmoving background stars) and a concentration around the 
origin (denoting comoving association members). We denote 
spectroscopically-confirmed members with filled circles and nonmembers 
with open circles, leaving only error bars for candidates without 
spectroscopy; we find general agreement between the two methods, with only 
one spectroscopic nonmember in the overall distribution of members. The 
black circle denotes our selection limit of $\sim$12 mas yr$^{-1}$; the 
spectroscopically-confirmed companion to RXJ1555.8-2512 appears to fall 
outside this limit, but its overall discrepancy (12.4 mas yr$^{-1}$ rounds 
down to the limit, so we consider it astrometrically confirmed. The 
overall agreement suggests that astrometric confirmation is typically 
sufficient for our purpose, though followup spectroscopy is very valuable 
for determining stellar properties and for avoiding the many systematic 
and astrophysical uncertainties of astrometry.}
 \end{figure*}

\subsection{Association Members and Background Stars}

\begin{deluxetable*}{llccclr}
\tabletypesize{\tiny}
\tablewidth{0pt}
\tablecaption{Status Determinations}
\tablehead{\colhead{Known Member} & \colhead{Candidate Companion} & 
\colhead{Spectroscopic} & \colhead{Astrometric} & \colhead{Final} & 
\colhead{Spectral} & \colhead{EW(H$\alpha$)}
\\
\colhead{} & \colhead{} & \colhead{Determination} & 
\colhead{Determination} & \colhead{Determination} & \colhead{Class}
}
\startdata
$\Delta$$K<$3&&&&&&\\
2M040807.82+280728.0&2M04080771+2807373&N&N?&N&K0 III&2.1\\
2M041618.85+275215.5&2M04161754+2751534&N&N?&N&G5 III&1.6\\
2M042134.60+270138.8&2M04213331+2701375&..&N?&N&..&..\\
2M044144.89+230151.3&2M04414565+2301580&Y&.&Y&M3&-5.7\\
CFHT-Tau-21&2M04221757+2654364&N&Y?&N&A5&4.8\\
CFHT-Tau-7&2M04321713+2421556&N&N?&N&F5&5.5\\
FW Tau&2M04292887+2616483&N&N?&N&G5 III&1.5\\
GM Aur&2M04551015+3021333&N&N?&N&K0 III&1.8\\
HBC 427&2M04560252+3020503&N&N?&N&K0 III&1.7\\
I04158+2805&2M04185906+2812456&N&..&N&G0 III&3\\
I04385+2550&2M04413842+2556448&..&N?&N&..&..\\
JH 112&2M04324938+2253082&Y&..&Y&M4.5&-22\\
LkCa 15&2M04391795+2221310&N&..&N&K3 III&1.3\\
V410 X-ray1&2M04175109+2829157&N&..&N&A0&9.8\\
V710 Tau AB&2M04315968+1821305&Y&Y?&Y&M3&-120\\
DENIS162041.5-242549.0&2M16204196-2426149&N&..&N&G-K V&1.6\\
GSC 06213-01459&GSC 06213-01459 B&..&Y?&Y&..&..\\
GSC 06784-00997&2M16101888-2502325&..&Y?&Y&..&..\\
GSC 06785-00476&2M15410726-2656254&..&Y?&Y&..&..\\
RXJ 1555.8-2512&2M15554839-2512174&Y&Y?&Y&M2.5&-6.6\\
RXJ 1558.8-2512&2M15585415-2512407&..&N?&N&..&..\\
RXJ 1602.8-2401B&2M16025116-2401502&..&Y?&Y&..&..\\
SCH160758.50-203948.90&2M16075693-2039424&..&N?&N&..&..\\
SCH160758.50-203948.90&2M16075796-2040087&Y&..&Y&$\sim$M1&..\\
SCH161511.15-242015.56&2M16151239-2420091&Y&Y?&Y&M4&-14.8\\
SCH161825.01-233810.68&2M16182365-2338268&N?&N?&N&K V&-6.9\\
SCH162135.91-235503.41&2M16213638-2355283&N&..&N&G-K V&..\\
ScoPMS042b&2M16102177-1904021&..&Y?&Y&..&..\\
ScoPMS048&ScoPMS 048 B&..&Y?&Y&..&..\\
UScoJ160245.4-193037&2M16024735-1930294&..&N?&N&..&..\\
UScoJ160700.1-203309&2M16065937-2033047&Y&Y?&Y&M2.5&-5.9\\
UScoJ160936.5-184800&2M16093658-1847409&..&N?&N&..&..\\
USco80&2M15583621-2348018&Y&Y?&Y&M3&-9.9\\
$\Delta$$K>$3&&&&&&\\
CFHT 4&2M04394921+2601479&N&..&N&K-M III&..\\
CoKu Tau/3&2M04354076+2411211&N&..&N&K-M V&..\\
DG Tau&2M04270370+2606067&..&N?&N&..&..\\
DO Tau&2M04382889+2611178&N&..&N&$<$K&..\\
FM Tau&2M04141556+2812484&N&..&N&K-M V&..\\
FO Tau&2M04144741+2812219&..&N?&N&..&..\\
I04216+2603&2M04244376+2610398&N&..&N&K-M V&..\\
IS Tau&2M04333746+2609550&N&..&N&K-M III&..\\
IS Tau&2M04333467+2609447&..&N?&N&..&..\\
LkCa 4&2M04162839+2807278&N&..&N&K-M III&..\\
MHO-Tau-2&2M04142440+2805596&N&..&N&$<$K&..\\
V410 X-ray 2&2M04183574+2830254&N&..&N&K-M V&..\\
V410 X-ray 5a&2M04190271+2822421&N&..&N&$<$K&..\\
X410 X-ray 6&2M04190223+2820039&N&..&N&K-M III&..\\
GSC 06784-00039&2M16084438-2602139&..&..&..&..&..\\
RXJ 1555.8-2512&2M15554788-2512172&..&N?&N&..&..\\
UScoJ161031.9-191305&2M16103232-1913085&..&Y?&Y&..&..\\
\enddata
\end{deluxetable*}

\begin{deluxetable*}{lllllr}
\tabletypesize{\tiny}
\tablewidth{0pt}
\tablecaption{Binary Properties}
\tablehead{\colhead{Primary} & \colhead{Secondary} & 
\colhead{$M_{prim}$} & \colhead{$M_{sec}$} & 
\colhead{$q$\tablenotemark{a}} & \colhead{$r$}
\\
\colhead{} & \colhead{} & \colhead{($M_{\sun}$)} & \colhead{($M_{\sun}$)} 
& \colhead{($M_s/M_p$)} & \colhead{(AU)}
}
\startdata
Known\\
2M04554757+3028077&2M04554801+3028050&0.20&0.14&0.70&915\\
DH Tau&DI Tau&0.64+0.044&0.64+(0.08)&1.06&2208\\
FS Tau&Haro 6-5B&0.64+0.33&0.82&0.85&2883\\
FV Tau&FV Tau/c&0.82+(0.62)&0.45+0.33&0.54&1782\\
FZ Tau&FY Tau&0.72&0.64&0.89&2490\\
GG Tau Aab&GG Tau Bab&0.72+0.60&0.14+0.044&0.14&1505\\
GK Tau&GI Tau&0.72+(0.027)&0.82&1.09&1905\\
HBC 352&HBC 353&2.26&0.94&0.42&1301\\
HBC 355&HBC 354&1.2&1.2&1.00&915\\
HN Tau A&HN Tau B&0.82&0.22&0.27&450\\
HP Tau-G2&HP Tau&2.49&0.94&0.38&3089\\
HP Tau-G2&HP Tau-G3&2.49&0.72+(0.10)&0.33&1463\\
HV Tau AB&HV Tau C&0.50+(0.31)&0.77&0.95&545\\
J1-4872 Aab&J1-4872 Bab&0.64+0.64&0.57+0.57&0.89&490\\
LkHa332-G1&LkHa332-G2&0.57+(0.57)&0.60+0.45&0.92&3753\\
MHO-Tau-1&MHO-Tau-2&0.45&0.45&1.00&570\\
UX Tau AC&UX Tau Bab&1.20+0.40&0.50+(0.40)&0.56&849\\
UZ Tau Aab&UZ Tau Bab&0.57+(0.16)&0.50+0.40&1.23&516\\
V710 Tau A&V710 Tau B&0.60&0.50&0.83&439\\
V773 Tau&2M04141188+2811535&1.20+0.94+0.60+(0.58)&0.09&0.027&3390\\
V807 Tau&GH Tau&0.82+0.50+(0.50)&0.50+0.50&0.55&3157\\
V928 Tau&CFHT-Tau-7&0.60+(0.60)&0.12&0.10&2646\\
V955 Tau&LkHa332-G2&0.74+0.45&0.60+0.45&0.88&1524\\
XZ Tau&HL Tau&0.50+0.33&0.82&0.99&3380\\
RXJ1558.1-2405A&RXJ1558.1-2405B&0.95+(0.14)&0.13+(0.03)&0.15&2632\\
RXJ1604.3-2130A&RXJ1604.3-2130B&1.12&0.49+(0.36)&0.76&2352\\
ScoPMS 052&RXJ1612.6-1859&1.35+0.49&0.60&0.33&2764\\
UScoJ160428.4-190441&UScoJ160428.0-19434&0.36+(0.36)&0.24&0.33&1417\\
UScoJ160611.9-193532 A&UScoJ160611.9-193532 B&0.13+0.13&0.13&0.50&1563\\
UScoJ160707.7-192715&UScoJ160708.7-192733&0.49+(0.08)&0.24&0.42&3400\\
UScoJ160822.4-193004&UScoJ160823.2-193001&0.60&0.68&1.13&1953\\
UScoJ160900.7-190852&UScoJ160900.0-190836&0.68&0.13&0.19&2743\\
UScoJ161010.4-194539&UScoJ161011.0-194603&0.36&0.13&0.36&3711\\
New\\
2M04414565+2301580 A\tablenotemark{b}&2M04414565+2301580 B&0.40&0.027&0.07&1794\\
JH112 A&JH112 B&0.72&0.22&0.31&951\\
V710 Tau AB&V710 Tau C&0.60+0.50&0.40&0.36&4056\\
GSC06213-01459 A&GSC06213-01459 B&0.87&(0.17)&0.19&461\\
GSC 06785-00476 A&GSC 06785-00476 B&1.51&(0.20)&0.13&914\\
GSC 06784-00997 A&GSC 06784-00997 B&0.60&(0.05)&0.09&697\\
RXJ1555.8-2512 A&RXJ1555.8-2512 B&1.65&0.43&0.26&1292\\
RXJ1602.8-2401B&RXJ1602.8-2401B&0.95&(0.11)&0.12&1047\\
2M16075796-2040087 A\tablenotemark{b}&2M16075796-2040087 B&0.7&0.074&0.10&3120\\
2M16151239-2420091 A\tablenotemark{b}&2M16151239-2420091 B&0.24&0.074&0.31&2604\\
ScoPMS042b A&ScoPMS042b B&0.36&(0.05)&0.14&664\\
ScoPMS048 A&ScoPMS048 B&1.12+0.24&(1.06)&0.78&442\\
2M16065937-2033047 A\tablenotemark{b}&2M16065937-2033047 B&0.49&0.43&0.88&1689\\
UScoJ161031.9-191305 A&UScoJ161031.9-191305 B&0.77&(0.033)&0.043&828\\
USco80 Aab\tablenotemark{b}&USco80 B&0.36+(0.36)&0.24&0.33&1779\\
\enddata
\tablecomments{Masses for all members with known spectral types were 
estimated using the mass-SpT relations described in Section 3.5, while 
masses in parentheses (for sources without spectral types) were 
estimated using the estimated mass of the system primary and the 
measured flux ratio. The references for these flux ratios are listed in 
Tables 1 and 3. Our model-dependent masses are uncertain to $\sim$20\%, 
and the mass ratios and projected separations have typical uncertainties 
of $\sim$10\%. Finally, some hierarchical multiple systems have mass 
ratios $q>$1, where the combined mass for all components of B is higher 
than that of A. We preserve the existing naming scheme for continuity, 
but will invert this mass ratio during our analysis (Section 5) to 
reflect that B is the most massive component.}
\tablenotetext{a}{For hierarchical multiple systems, we computed the mass 
ratio by summing the individual stellar masses in all sub-components of 
the wide "primary" and "secondary".}
\tablenotetext{b}{Several newly-identified companions appear to be more 
massive than the known member, suggesting that the known member is the 
binary secondary. In cases where the known member had a generic name 
(i.e. USco80), we have appropriated that name for the new 
member to avoid name proliferation in the literature. For systems with 
coordinate-based names, we have used the 2MASS name of the new 
member to avoid confusion over coordinates.}
\end{deluxetable*}

In Table 8, we summarize our spectroscopic and astrometric membership 
assessments for each candidate young stars in our sample, along with the 
final membership assessments that we will use in our subsequent 
statistical arguments. We found that 11 of the 18 USco candidates and 3 of 
the 15 Taurus candidates with separations of 3-30$\arcsec$\, and flux 
ratios $\Delta$$K\la$3 are comoving young stars, while most of the 
candidates that we considered with more extreme flux ratios are not 
associated. We were not able to test the association of one USco 
candidates with a larger flux ratio, and even though another appears 
comoving, its faintness and the high density of stars in the direction of 
Upper Sco (and thus the bulge) suggests that cutting our statistical 
analysis at $\Delta$$K\la$3 would be prudent.

The total number of confirmed background stars (28 in Taurus and 9 in 
Upper Sco) is consistent within $\la$2$\sigma$ with the number that we 
projected in our original survey (36$\pm$6 and 16$\pm$4). In Table 9, we 
list the stellar properties for each pair of newly-confirmed young stars, 
plus all of the pairs listed in Table 3; we derived these properties using 
the methods described in Section 3.4. The mass ratios for hierarchical 
triple systems were computed by summing all sub-components within each 
member of the wide pair.

\section{The Properties of Wide Binary Systems}

\begin{figure*}
\epsscale{1.00}
\plotone{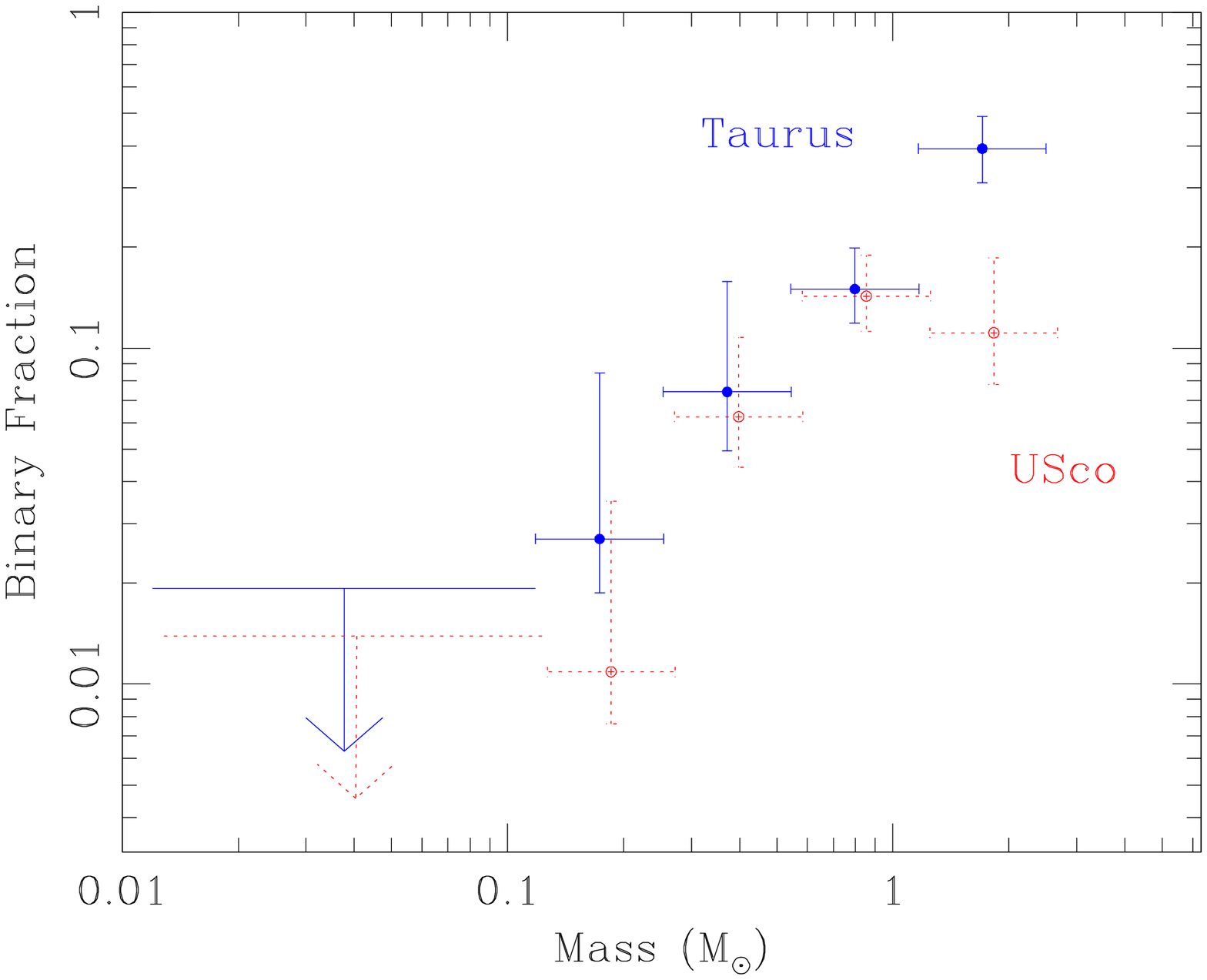}
\caption{Wide binary frequency as a function of primary mass. The overall 
binary frequency declines with mass, reaching upper limits of $\sim$1-2\% 
for the substellar regime ($M\la$0.1 $M_{\sun}$). The binary frequency for 
high-mass stars (1.15-2.50 $M_{\sun}$) is significantly higher in Taurus 
than in Upper Sco, but otherwise, the binary frequencies are not 
significantly different.} 
\end{figure*}

\begin{figure*}
\epsscale{1.00}
\plotone{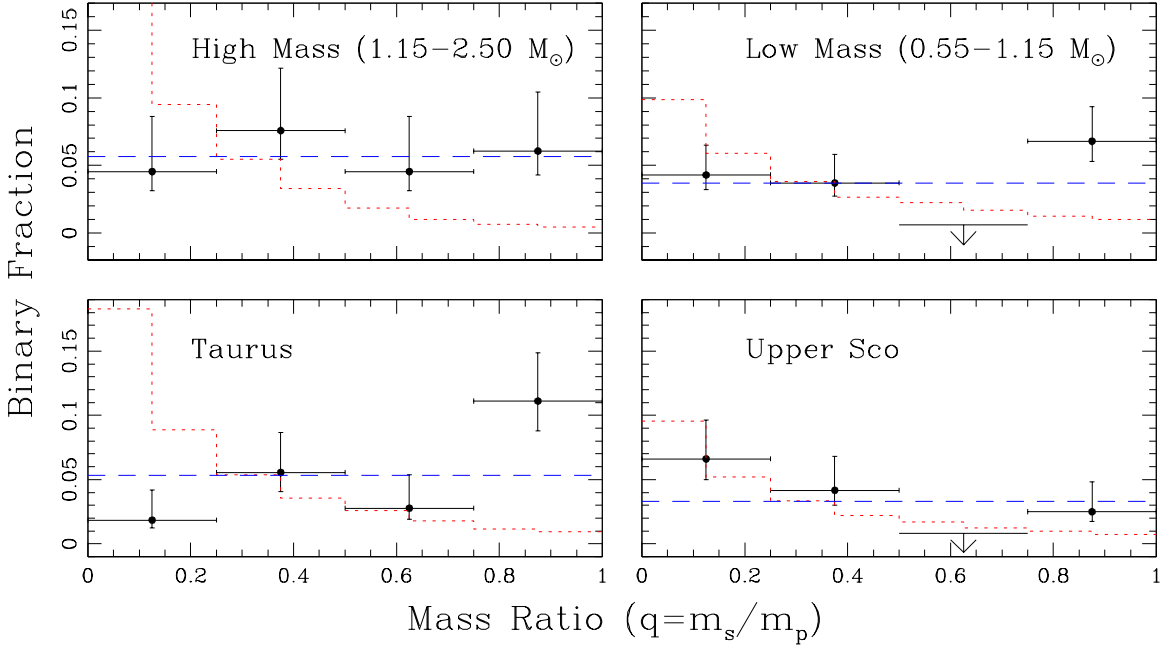}
\caption{Top: Mass ratio distribution for high-mass stars (1.15-2.50 
$M_{\sun}$) and intermediate-mass stars (0.55-1.15 $M_{\sun}$). Bottom: 
Mass ratio distribution for Taurus and Upper Sco when the two mass ranges 
are combined (0.55-2.50 $M_{\sun}$). The lowest bin is incomplete at 
$q$$\la$0.02-0.04, but this should not affect our results 
because companions with such extreme mass ratios do not seem to form 
often (e.g. Kraus et al. 2008). In each case, we also plot the expected 
distribution if the companions were drawn randomly from an IMF (red 
dotted line) or from a constant distribution (blue dashed line) with 
the same frequency. The IMF does not produce a satisfactory fit for most 
cases, but a constant distribution does. Finally, we also note that the 
shape of the IMF distribution varies between subsamples, depending on 
the masses of the primary stars that make up those subsamples. The first 
IMF bin for the high-mass subsample is 24\%; we truncated the plot at 
17\% in order to improve resolution for the other bins and the 
intermediate-mass subsample.}
\end{figure*}

\begin{figure*}
\epsscale{1.00}
\plotone{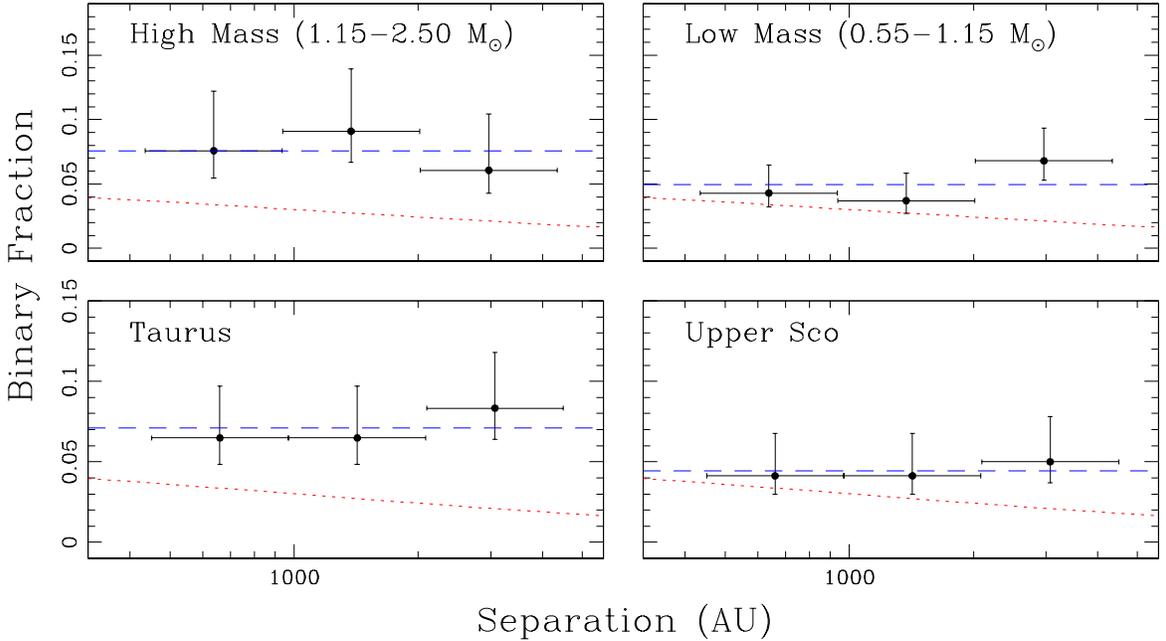}
\caption{Separation distributions for the same four subsamples plotted 
in Figure 7. We also plot the log-normal separation distribution found 
by DM91 for field solar-mass stars, normalized to the DM91 binary 
frequency (red dotted line), and a log-constant distribution normalized 
to the same binary frequency as that subsample (blue dashed line). The 
DM91 distribution underpredicts the overall binary frequency for 
high-mass stars and Taurus, and even the expected trend (declining 
frequency with increasing separation) does not match with the data. The 
log-constant distribution produces a better fit in all cases. Even if we 
renormalize the DM91 function to our binary frequency, it still does not 
fit our intermediate-mass or Upper Sco subsamples.}
\end{figure*}

\begin{figure*}
\epsscale{1.00}
\plotone{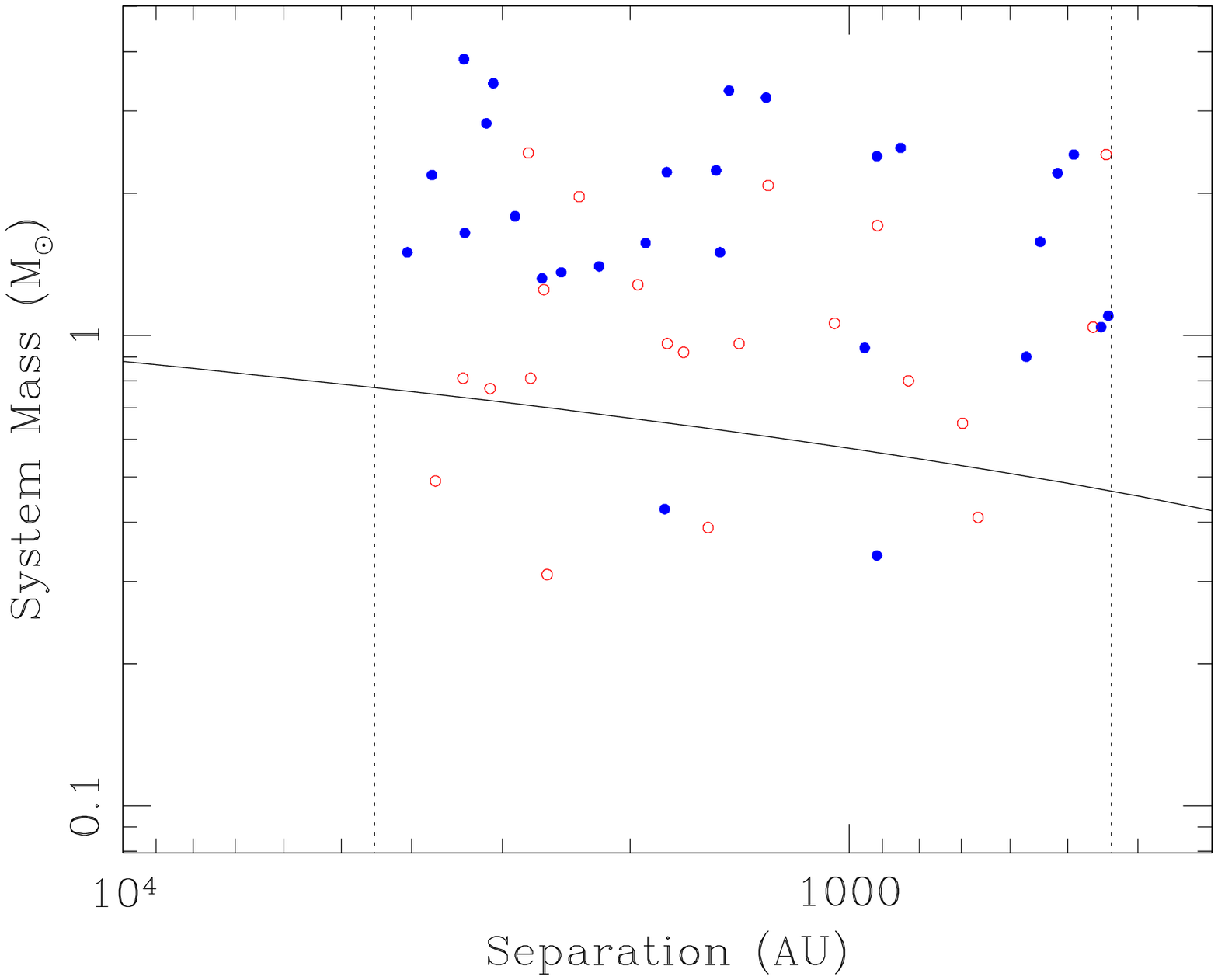}
\caption{Total system mass as a function of separation for all of our 
wide binary systems in Taurus (blue filled circles) and Upper Sco (red 
open circles). We also show the empirical ``maximum separation limit'' 
observed in the field by Reid et al. (2001) and Burgasser et al. (2003) 
(solid line) and the separation limits of our survey (dotted lines). 
Six pairs with masses of $\ga$0.3-0.4 $M_{\sun}$ exceed the empirical 
mass-separation limit, suggesting that it might not be a primordial 
feature for these higher-mass systems. However, we found no wide binary 
systems with total masses of $\la$0.3 $M_{\sun}$, suggesting that there is 
a genuine primordial paucity of wide low-mass systems.}
\end{figure*}

In the following subsections, we explore the implications of our survey of 
wide (500--5000) multiplicity. In Section 5.1, we examine the 
mass-dependent frequency of wide binary systems for each association and 
discuss the differences between Taurus and Upper Sco. In Sections 5.2 and 
5.3, we examine the mass ratio distributions and separation distributions 
for each association and in two different mass ranges, then compare them 
to functional forms that might be expected. Finally, in Section 5.4, we 
examine the separation as a function of mass for our new binary systems 
and compare our sample to the empirical upper limit that has been 
suggested based on field multiplicity surveys.

As we described in our preliminary survey (KH07a, Section 3.3 and Figure 
2) and in Section 4, our census of this separation range is complete for 
all candidate companions brighter than $K=14.3$ ($\sim$15 $M_{Jup}$ in 
Taurus or $\sim$20 $M_{Jup}$ in Upper Sco), except for two candidate 
companions in Upper Sco with $\Delta$$K\ga$3.75 ($q<0.05$, if they are 
associated) that we were not able to observe. Our survey also could not 
reach fainter than $\Delta$$K\sim$5.5 at separations of 3-5\arcsec\, so it 
is possible that some close candidate companions with extreme mass ratios 
might have been missed around the highest-mass stars. However, there is 
only one such companion at separations $>$5\arcsec\, in Taurus 
(2M04141188+2811535), which suggests that the probability is low. We note 
that there is one triple system (the nonhierarchical HP Tau-G2, HP Tau, 
and HP Tau-G3) where all three components fall in this separation range; 
we will treat HP Tau and HP Tau-G3 as independent companions to HP Tau-G2 
for statistical purposes. There is also a probable triple system (the 
possibly hierarchical V955 Tau, LkHa332-G1, and LkHa332-G2) where 
LkHa332-G2 is $\sim$11\arcsec\, away from V955 Tau and $\sim$26\arcsec\, 
away from LkHa332-G1, but V955 Tau and LkHa332-G1 are $>$30\arcsec\, 
apart. Since all three of these objects have very similar masses 
(1.05-1.20 $M_{\sun}$, all being close binary pairs) and it's not clear if 
the system is truly hierarchical, we will consider this triplet as a 
closer 11\arcsec\, pair and a wider 26\arcsec\, pair. Finally, for all 
hierarchical systems, we have treated each component of the wide pair as a 
single object with the summed mass of all sub-components.

\subsection{The Mass Dependence of the Wide Binary Frequency}

Field surveys have shown that the binary frequency and binary separation 
distribution both decline with decreasing mass, implying that the wide 
($\sim$500-5000 AU) binary frequency should strongly decline over the 
mass range of our sample. Our preliminary survey paper (KH07a) also 
found this trend at young ages, suggesting that it is a primordial 
effect. However, we also found the wide binary frequency for a given 
mass to be higher in the lowest-density regions, like Taurus and 
Chamaeleon-I, than in moderately denser regions like Upper Sco.

In Figure 6, we plot the mass-dependent binary frequency for four sets 
of masses in the stellar regime, plus all sources near and below the 
substellar boundary. The complete sample comprises all of the stars that 
we considered in our original survey (KH07a), with all confirmed binary 
systems drawn from Table 9 of this work. In both associations, the 
binary frequency clearly declines over the full mass range; we found 
frequencies of $\ga$10\% for stars more massive than $\sim$1 $M_{\sun}$, 
declining to upper limits of $\la$1-2\% in the substellar regime. This 
decline appears to be relatively smooth and monotonic in Taurus, but it 
is unclear whether Upper Sco features a shallower version of the decline 
or a more abrupt shift from a high value to a low value at $\sim$0.5 
$M_{\sun}$.

The binary frequency is similar across most of the mass range for these 
two associations. This result differs from our initial statistical 
sample, but adding additional systems with larger separations or mass 
ratios drove the two distributions closer together in our updated 
analysis. However, we have again found a significantly higher binary 
frequency among the highest-mass stars in Taurus as compared to their 
brethren in Upper Sco; this result was the only highly significant 
difference in our initial analysis, and our updated results find it to 
be a 4$\sigma$ effect.

This regional difference among the higher-mass stars in our sample is 
difficult to explain in terms of binary destruction processes. Dynamical 
disruption (perhaps due to a more crowded natal environment) should 
preferentially destroy low-mass binaries before high-mass binaries. The 
similarity between the two environments in the lower-mass regime seems to 
rule disruption out. However, observations of mass segregation at very 
young ages (e.g. Hillenbrand \& Hartmann 1998; Sirianni et al. 2002) 
indicate that perhaps stars might be primordially mass-segregated, with 
higher-mass stars forming preferentially in denser parts of their natal 
environment. Binary disruption in these denser regions should be 
significantly enhanced as compared to the sparse outer reaches of a 
collapsing molecular cloud. High-mass stars are significantly less common 
than their lower-mass counterparts, so even if these dense central areas 
also caused the disruption of lower-mass binaries, it might not be 
strongly reflected in the overall binary population (which could be 
dominated by a majority of systems that form outside the densest 
concentrations).

\subsection{The Mass Ratio Distribution of Wide Binaries}

Field surveys have also suggested that the mass ratio distribution 
varies significantly with primary mass. DM91 found that G dwarfs 
tend to have lower mass companions (with a modal mass ratio of 
$q\sim$0.3), while surveys of M dwarfs by FM92 and RG97 found a 
flat distribution and several recent surveys of brown dwarfs (e.g. 
Close et al. 2003; Burgasser et al. 2003; Bouy et al. 2003) found 
that their mass ratios are sharply peaked toward unity. By 
contrast, surveys of young associations have found that flat mass 
ratio distributions seem to dominate across a range of system 
masses, from $\sim$2 $M_{\sun}$ to at least as low as 0.5 
$M_{\sun}$ (e.g. Kraus et al. 2008 for Upper Sco), though a 
distribution biased toward unity seems to be universal among the 
lowest-mass stars and brown dwarfs (Kraus et al. 2006; Ahmic et 
al. 2007).

In Figure 7, we plot the mass ratio distribution for four subsets 
of our sample. In the top panels, we show the distribution 
spanning both associations for the highest-mass bin (1.15-2.50 
$M_{\sun}$) as compared to the intermediate-mass bin (0.55-1.15 
$M_{\sun}$), while in the bottom panels, we show the distribution 
for both mass bins as determined individually in Taurus and Upper 
Sco. We also show two possible mass ratio distributions: a flat 
distribution and a distribution where companions are randomly 
drawn from the IMF. We adopted our IMF from the spectroscopic 
membership surveys of Upper Sco by Preibisch et al. (1998, 2002) 
and Slesnick et al. (2006a); this function is defined as a broken 
power law (Scalo 1998; Kroupa 2002): 
$\Psi$$(M)=dN/dM$$\propto$$M^{-\alpha}$, where $\alpha=-2.8$ for 
$0.6<M<2.5$ $M_{\sun}$, $\alpha=-0.9$ for $0.15<M<0.6$ $M_{\sun}$, 
and $\alpha=-0.6$ for $0.02<M<0.15$ $M_{\sun}$. This broken power 
law mass function is rougly equivalent to the continuous 
log-normal mass function that has also been suggested (Miller \& 
Scalo 1979; Chabrier 2001). Several other possible mass ratio 
distributions have been suggested, including a truncated Gaussian 
(DM91) and a log-normal distribution (Kraus et al. 2008), but the 
first has been largely discounted by now and the latter does not 
differ significantly from a flat distribution given our sample 
size. The wider array of possible mass ratio distributions has 
been summarized and weighed by Kouwenhoven et al. (2009), but our 
sample size does not allow most of the fine distinctions found in 
that paper.

We have found that drawing companions from the IMF produces a very poor 
fit in most cases; a one-sample Kolmogorov-Smirnov test finds a normalized 
maximum difference $D$ (with respect to the model) of $D_H=0.53$ for the 
high-mass subset (1.15-2.50 $M_{\sun}$), $D_L=0.40$ for the 
intermediate-mass subset (0.55-1.15 $M_{\sun}$), $D_T=0.57$ for the Taurus 
subset, and $D_U=0.21$ for the Upper Sco subset. The first three results 
all imply disagreement at $P\ga$99\%, but the Upper Sco subset (which is 
smallest, $N=16$) is not inconsistent ($P<$80\%). The flat distribution 
yields $D_H=0.17$, $D_L=0.27$, $D_T=0.33$, and $D_U=0.42$, respectively, 
or confidence values of $P<$80\%, $P\sim$90\%, $P\sim$99\%, and 
$P\sim$99\%. The goodness of fit for the Upper Sco subsample is 
significantly worse than for the IMF-derived distribution, but the others 
all have better goodness of fit (though the low-mass and Taurus results 
still indicate disagreement).

Our results for Taurus and for both mass ranges are similiar to 
those that we reported for close binaries in Upper Sco (Kraus et 
al. 2008), with similar-mass companions typically over-represented 
compared to the IMF. Our results for wide binaries in Upper Sco 
show little evidence of this trend, but the sample is also smaller 
than for Taurus. We also note that among the low-mass subsample, 
Taurus binaries have predominantly similar masses (9/12 with 
$q>$0.75) while Upper Sco binaries tend to have low-mass 
secondaries (6/11 with $q<$0.25). Dividing the sample this finely 
reduces the significance of our results even further, especially 
since most of the solar-type stars in Upper Sco remain 
unidentified and the current census could be subject to some 
unknown bias, but this difference in the mass ratio distributions 
presents an intriguing hint of an environmental effect. As a 
whole, though, our results argue against a mechanism that forms 
binaries via random pairing, including their formation in entirely 
separate cloud cores. Our results also suggest that the masses of 
binary companions could be selected via a similar process across a 
wide range of separations, given that the mass ratio distribution 
is mostly similar at separations spanning 5 to 5000 AU.

Finally, we note that this distribution could be replicated by 
forming wide binaries out of small-N clusters, since dynamical 
interactions could force out the lower-mass members and leave the 
two highest-mass members as a bound pair. However, other features 
of pre-main sequence stars place strict limits on the amount of 
dynamical sculpting in these early groups. Most young stars in 
this mass range have disks at ages of 1-2 Myr, including many wide 
binary components (e.g. Furlan et al. 2006; Scholz et al. 2006), 
which suggests that they have not been involved in any energetic 
interactions. Also, many lower-mass stars ($M$$\sim$0.4-0.7 
$M_{\sun}$) are found in binaries with separations of 10-500 AU 
(e.g. Kraus et al. 2008), and few such binaries would survive in a 
dynamically active environment. These observations seem to suggest 
that a dynamical solution can not simultaneously satisfy all of 
the data.

\subsection{The Separation Distribution of Wide Binaries}

Finally, the binary parameter that varies most distinctly among field 
systems is the separation distribution. DM91 found that G dwarfs have a 
mean separation of $\sim$30 AU and some systems are as wide as 
$\sim$10$^4$ AU, while the recent substellar surveys have found a mean 
separation of $\sim$4 AU and very few systems wider than 20 AU, and the 
M dwarf surveys of FM92 and RG97 seem to suggest intermediate 
properties. Our results for smaller separations in Upper Sco (Kraus et 
al. 2008) are not strongly indicative because that survey spanned the 
peak of the DM91 distribution (where it is approximately flat in 
log-separation), but it appears that there is no significant difference 
in the separation distribution between 0.5 and 2 $M_{\sun}$ across a 
range of 5-500 AU. In Figure 8, we plot the separation distribution of 
our sample of wide binary systems, spanning separations of 500-5000 AU, 
as well as the separation distribution suggested by DM91 (a log-normal 
function) and a log-constant distribution. As for Figure 7, we compare 
our high-mass and intermediate-mass samples (top) and our Taurus and 
Upper Sco samples (bottom).

In all cases, it appears that the companion frequency increases or is flat 
with increasing separation. When we test the log-constant distribution 
with a one-sample Kolmogorov-Smirnov test, we typically find good 
agreement with normalized maximum cumulative differences of $D_H=0.16$, 
$D_L=0.19$, $D_T=0.14$, and $D_U=0.11$ for the high-mass, low-mass, 
Taurus, and Upper Sco subsets. In all cases, the confidence level is 
$<$85\%. This is not unexpected; our results for two-point correlation 
functions indicate that the separation distribution function is 
approximately log-flat out to even larger separations ($\sim$20,000 AU; 
Kraus \& Hillenbrand 2008). Kouwenhoven et al. (2007) also reported that 
the log-flat separation distribution produces a satisfactory fit for 
higher-mass binaries in Sco-Cen.

When we test the DM91 separation distribution with a one-sample 
Kolmogorov-Smirnov test, we find results that are less consistent, but not 
necessarily inconsistent: $D_H=0.23$, $D_L=0.28$, $D_T=0.22$, and 
$D_U=0.19$. The high-mass, Taurus, and Upper Sco subsamples are not 
inconsistent ($P<$85\%), but the low-mass sample disagrees at $P\sim$96\%. 
However, given our results for two-point correlation functions that 
support the log-flat distribution conclusively at larger separations, we 
find it preferable to the log-normal distribution. We also note that the 
DM91 distribution is independently normalized by the DM91 binary 
frequency, and K-S tests ignore the binary frequencies by implicitly 
renormalizing them to the same value. It is illustrative to preserve this 
normalization by using a $\chi$$^2$ test. We found fit parameters of 
$\chi$$^2_H=15.7$ for the high-mass subset, $\chi$$^2_L=11.6$ for the 
low-mass subset, $\chi$$^2_T=19.2$ for the Taurus subset, and 
$\chi$$^2_U=6.5$ for the Upper Sco subset. The high-mass and Taurus 
subsets disagree at very high confidence ($\ga$99.9\%, while the low-mass 
subset disagrees at $P\sim$99\% and the Upper Sco subset disagrees at 
$P\sim$90\%. We therefore confirm the well-known result that the DM91 
binary frequency is less than the binary frequency for these young stellar 
populations, indicating that binary companions are over-abundant with 
respect to the field (e.g. Ghez et al. 1994; Kouwenhoven et al. 2007).

The presence of a log-flat primordial separation distribution suggests that 
the field separation distribution may be a result of post-natal dynamical 
evolution. The stars in these associations should escape to the field with 
no further sculpting, and the dynamical simulations of Weinberg et al. 
(1987) suggest that the field stellar density is too low to affect 
binaries closer than $\sim$$10^4$ AU. However, it has been suggested that 
many (or perhaps even most) stars are born in much denser clusters (Lada 
\& Lada 2003), though there are also arguments to the contrary (Adams \& 
Myers 2001). If this model is true, then the majority of stars could 
linger in a relatively high-density environment for up to several Gyr. 
Observations suggest that the cluster environment is typically dense 
enough to remove most of the binaries with separations of $\ga$100 AU 
(e.g. Praesepe, Patience et al. 2002; Coma Ber, Kraus et al., in prep).

Therefore, the field population almost certainly represents a mix 
of binary populations, a suggestion discussed by Kroupa (1998) and 
Kroupa et al. (1999). Those stars which are born in T associations 
and OB associations enter the field almost immediately, with their 
wide binary population nearly intact. In contrast, stars that form 
in clusters are stripped of their outer binary companions, with 
the degree of stripping depending on the density of the cluster 
environment, the density evolution over time, and the elapsed time 
until a typical star is tidally removed and joins the field 
(Kroupa et al. 2001; Kroupa \& Bouvier 2003). A survey of wide 
binary systems in young clusters like the ONC or IC348 should 
directly reveal this sculpting process, but the crowded 
environment makes it difficult to distinguish bound binary systems 
from chance alignments (e.g. Simon 1997; K\"ohler et al. 2006).

We must add a caveat that the primordial multiplicity of dense 
clusters is still not well-constrained for wide separations, 
especially at $\ga$500 AU where it is impossible to distinguish 
bound companions from chance alignments. The absence of wide 
binary systems in open clusters does not necessarily indicate that 
they form and are disrupted; a primordial deficiency of wide 
binary systems could also explain the data. Studies of the ONC by 
K\"ohler et al. (2006) and Reipurth et al. (2007) find that the 
binary frequency at smaller separations ($\sim$60--600 AU) is a 
factor of $\ga$2 lower than in Taurus-Auriga and Sco-Cen, though 
low number statistics forced their measurements to span a wide 
range of primary masses that might not be equally represented in 
the surveys of closer associations. K\"ohler et al. further 
suggest that there is little evidence of a density dependence 
between the core and halo of the ONC, arguing against a dynamical 
origin of the lower binary frequency. However, the larger sample 
studied by Reipurth et al. shows a steep decrease in the 
separation distribution at $\sim$225 AU that is most pronounced in 
the cluster core, indicating a possible signature of dynamical 
disruption for wider binary systems. In addition, both of these 
results depend on the membership census of the ONC (e.g. Jones \& 
Walker 1988; Hillenbrand 1997), which is still uncertain for many 
candidates.

The most compelling argument for an environmental difference in 
the primordial binary properties was set forth by Durisen \& 
Sterzik (1994) and Sterzik et al. (2003), who predicted that 
regions with a higher gas temperature should have a binary 
separation distribution that is biased to smaller values. One 
source of this heating could be nearby high-mass stars, which 
would naturally predict the absence of high-mass binary systems in 
dense clusters with numerous OB stars. However, feedback from 
these high-mass stars should dispel the natal gas and shut down 
star formation, so delicate timing would be required in order for 
this effect to play a significant role. An indirect test of the 
primordial binary properties was attempted by Kroupa et al. (1999) 
by using N-body simulations to evolve several candidate proto-ONC 
clusters forward to the present day. They concluded that in order 
to fit the current dynamical state, a binary frequency lower than 
in Taurus-Auriga was required. However, they only tested six model 
populations, so their simulation results could include significant 
degeneracy between choices of parameters. There have also been 
numerous observational advances in the past decade, and the 
simulated results of Kroupa et al. should be confronted with these 
new findings.

Finally, if the separation distribution is truly log-flat for Taurus and 
Upper Sco, then there is at most a moderate decrement with respect to the 
binary separation distribution at smaller separations. Our previous 
high-resolution imaging survey of Upper Sco (Kraus et al. 2008) found that 
for separations of 5-500 AU and primary masses of 0.5-2.0 $M_{\sun}$, the 
binary frequency is 19$^{+3}_{-2}$\% per decade of separation. In our wide 
binary sample spanning 500-5000 AU, the corresponding frequencies are 
23$^{+6}_{-4}$\% for the high-mass subsample, 15$^{+3}_{-2}$\% for the 
low-mass subsample, 21$^{+4}_{-3}$\% for the Taurus subsample, and 
13$^{+4}_{-3}$\% for the Upper Sco subsample. A comprehensive multiplicity 
survey of Taurus will be required to place these statistics in context, 
but we find it intriguing that the binary frequency is so similar across 
three decades of separation (or 9 decades of mean density in the original 
cloud core). Either a single binary formation process operates across the 
full range of length scales, or several binary formation processes all 
yield similar frequencies.

\subsection{Unusually Wide Binary Systems}

As we described above, the separation distribution in the field seems to 
be strongly mass-dependent. Field surveys also suggest an empirical 
relation between the total mass of a system and its maximum possible 
separation, where the relation is logarithmic in the solar-mass regime 
($\log{a_{max}}=3.3M_{tot}+1.1$ if $M_{tot}$$\ga$0.3 $M_{\sun}$; Reid et 
al. 2001) and quadratic in the low-mass regime ($a_{max}=1400M_{tot}^2$ 
if $M_{tot}$$\la$0.3 $M_{\sun}$; Burgasser et al. 2003). This relation 
also provides a good working definition for what might be considered an 
``unusually wide'' binary system; many such systems have been reported 
in nearby star-forming regions, but the absence of a rigorous definition 
has led to much confusion regarding their true uniqueness.

Our results suggest that the binary frequency is strongly mass-dependent 
for young stars, but the form of the separation distribution may not 
change significantly. If the field $a_{max}$-$M_{tot}$ relation is 
genuinely primordial, then our separation-limited (500-5000 AU) sample 
should include no binary systems with masses of $M_{tot}$$\la$0.5 
$M_{\sun}$ and a limited range of separations for 
0.5$\la$$M_{tot}$$\la$0.8 $M_{\sun}$. However, if the field star 
population (which mostly forms in clusters) is sculpted by post-natal 
dynamical interactions in those clusters, then these limits might not be 
present in our sample.

In Figure 9, we plot the projected separation and total mass of each of 
the systems in our survey, plus the empirical $a_{max}$-$M_{tot}$ 
relation observed in the field. As we noted in the previous section, 
there is a genuine paucity of wide systems among the lowest-mass 
members, so any additional systems discovered with $M_{tot}\la$0.3 
$M_{\sun}$ should be considered genuinely ``unusual''. However, we see 
six intermediate-mass systems that seem to exceed this limit, and no 
evidence of an outer envelope. Our sample includes six systems that all 
have a total mass of $\sim$0.3-0.4 $M_{\sun}$, and even they seem to 
span the full separation range of our survey. Some of these systems 
could be chance alignments of two low-mass stars, but this number must 
be small because there are none among the least-massive third of our 
sample ($M\la$0.3 $M_{\sun}$). Based on our analysis of the associations' 
two-point correlation functions (Kraus \& Hillenbrand 2008), we expect 
$\la$2 chance alignments in Upper Sco and $\la$1 chance alignments in 
Taurus for all unassociated pairs of members with $M<$0.4 $M_{\sun}$, 
whereas we actually observe four and two, respectively. We would 
also expect chance alignments to be concentrated at the largest 
separations, not distributed evenly in logarithmic separation, and to 
include more pairs with a total mass $<$0.3 $M_{\sun}$.

Our survey shows that in a dynamically unevolved population like Taurus 
or Upper Sco, 6$^{+3}_{-2}$\% (Taurus 2/31, USco 4/65) of all single stars 
or binary systems with a total mass of 0.25$<M<$0.50 $M_{\sun}$ have a 
companion with a projected separation of 500--5000 AU. As a result, at 
least this many systems exceed the field $M_{tot}$-$a_{max}$ limit. By 
contrast, $<$0.4\% (Taurus 0/89, USco 0/167) of all binary systems 
or single stars with a total mass of $<$0.25 $M_{sun}$ have such a wide 
companion. The first result implies that the field $M_{tot}$-$a_{max}$ 
relation is another consequence of dynamical sculpting for the majority of 
field stars that form in dense clusters. Systems with lower binding energy 
are more prone to disruption in a dense environment, so high-mass systems 
can maintain wider binary components than their lower-mass brethren. 
However, dynamical sculpting can not explain the sharp paucity of 
primordial wide systems below $M_{tot}\sim$0.3 $M_{\sun}$, or that wide 
systems seem to decline rapidly in frequency below $M_{tot}\sim$0.7-0.8 
$M_{\sun}$. This result could indicate a critical mass limit for 
large-scale fragmentation of a collapsing cloud core.

\section{Summary}

In this paper, we have presented an astrometric and spectroscopic followup 
campaign to confirm the youth and association of a complete sample of wide 
binary companions to intermediate- and low-mass stars 
(2.5$>$$M_{prim}$$>$0.02 $M_{\sun}$. Our survey found fifteen new wide 
binary companions with separations of 3-30\arcsec\, ($\sim$500--5000 AU), 
3 in Taurus and 12 in Upper Sco, raising the total number of such systems 
to 49. Our survey should be complete for all companions with masses 
$M_{sec}$$\ga$15-20 $M_{Jup}$ and mass ratios $q\ga$0.02-0.04.

In some respects, this wide binary population conforms to expectations 
from field multiplicity surveys; higher-mass stars have a higher 
frequency of wide binary companions, and there is a marked paucity of 
wide binary systems near and below the substellar regime. However, this 
wide binary population also deviates significantly from other 
established properties of field binary systems. The separation 
distribution appears to be nearly log-flat across a very wide range of 
separations (5-5000 AU), and the mass ratio distribution seems more 
biased toward similar-mass companions than would be expected for an 
IMF-shaped distribution or from the field G-dwarf distribution. Finally, 
the maximum binary separation also shows markedly different behavior, 
with no evidence of a mass-dependent separation limit for system masses 
$\ga$0.3 $M_{\sun}$ and abrupt cessation of any wide binary formation 
(for separations $\ga$500 AU) below this limit.

We attribute these differences to the post-natal dynamical sculpting 
that occurs for most field systems. All of the systems in our sample, 
which come from unbound low-density associations, will escape to the 
field without further dynamical evolution. However, most stars seem to 
form in denser clusters; even if a wide binary population forms for 
these stars, it will most likely be stripped before the stars can escape 
into the field. This explanation suggests that the properties of wide 
binary systems in the field are not representative of their formation 
process.

Finally, we note that wide ($\sim$500-5000 AU) binary systems with total 
masses of $\la$0.3 $M_{\sun}$ appear to be very rare at all ages, 
suggesting that any system in this range of parameter space is indeed 
``unusually wide''. However, additional followup is required to 
determine the true total mass of a system, as there are many 
hierarchical multiple systems (e.g. USco80 and UScoJ160611.9-193533) 
that could masquerade as ``unusually wide low-mass binaries'' until AO 
and radial velocity surveys discover their higher-order multiplicity.

\acknowledgements

We thank Russel White and Michael Ireland for several helpful discussions 
regarding wide binary formation, Cathy Slesnick and Greg Herczeg for 
discussions regarding the interpretation of young star spectra, and Brian 
Cameron and Stan Metchev for discussions of astrometric precision and 
accuracy in AO imaging. We also thank the referee for a prompt and 
insightful report. Finally, we thank the Keck LGSAO team for their efforts 
in developing and supporting a valuable addition to the observatory. ALK 
was supported by a NASA Origins grant to LAH. This work makes use of data 
products from 2MASS, which is a joint project of the University of 
Massachusetts and the IPAC/Caltech, funded by NASA and the NSF. This work 
also makes use of data products from the DENIS project, which has been 
partly funded by the SCIENCE and the HCM plans of the European Commission 
under grants CT920791 and CT940627. Finally, our research has made use of 
the USNOFS Image and Catalogue Archive operated by the USNO, Flagstaff 
Station (http://www.nofs.navy.mil/data/fchpix/).

Some of the data presented herein were obtained at the W.M. Keck 
Observatory, which is operated as a scientific partnership between 
Caltech, the University of California, and NASA. The observatory was made 
possible by the generous financial support of the W.M. Keck Foundation. 
The authors also wish to recognize and acknowledge the very significant 
cultural role and reverence that the summit of Mauna Kea has always had 
within the indigenous Hawaiian community. We are most fortunate to have 
the opportunity to conduct observations from this mountain.

\facility{Hale, KeckII, CTIO:2MASS, FLWO:2MASS}


\begin{thebibliography}{}
\bibitem[Adams \& Myers(2001)]{am01} Adams, F. \& Myers, P. 2001, \apj, 
553, 744
\bibitem[Ahmic et al.(2007)]{ahm07} Ahmic, M., Jayawardhana, R.,
Brandeker, A., Scholz, A., van Kerkwijk, M., Delgado-Donate, E., \&
Froebrich, D. 2007, \apj, 671, 2074
\bibitem[Allen \& Strom(1995)]{as95} Allen, L. \& Strom, S. 1995, \aj, 109, 1379
\bibitem[Ardila et al.(2000)]{ard00} Ardila, D., Mart\'in, E., \& Basri, G. 
2000, \aj, 120, 479
\bibitem[Baraffe et al.(1998)]{bcah98} Baraffe, I., Chabrier, G., Allard, F. \& 
Hauschildt, P.H. 1998, \aap, 337, 403
\bibitem[Bate et al.(1998)]{bate98} Bate, M., Clarke, C., \& McCaughrean, M. 1998, \mnras, 297, 1163
\bibitem[Bessell \& Brett(1988)]{bb88} Bessell, M. \& Brett, J. 1988,\pasp, 100, 1134
\bibitem[Boden et al.(2007)]{bod07} Boden, A. et al. 2007, \apj, 670, 1214
\bibitem[Bonnarel et al.(2000)]{bon00} Bonnarel, F. et al. 2000, \aaps,143, 33
\bibitem[Bouy et al.(2003)]{bouy03} Bouy, H., Brandner, W., Mart\'in, E.,
Delfosse, X., Allard, F., \& Basri, G. 2003, \aj, 126, 1526
\bibitem[Bouy et al.(2007)]{bouy07} Bouy, H., Huelamo, N., Mart\'in, E., 
Barrado y Navascues, D., Sterzik, M., \& Pantin, E. 2007, \aap, 463, 641
\bibitem[Briceno et al.(1998)]{b98} Briceno, C., Hartmann, L., Stauffer, J., \& 
Mart\'in, E. 1998, \aj, 115, 2074
\bibitem[Burgasser et al.(2003)]{burg03} Burgasser, A., Kickpatrick, J.D., Reid, I.N., Brown, M., 
Miskey, C., \& Gizis, J. 2003, \apj, 125, 850
\bibitem[Caballero et al.(2006)]{c06} Caballero, J., Mart\'in, E., Dobbie,
P., \& Barrado y Navascues, D. 2006, \aap, 460, 635
\bibitem[Cameron et al.(2009)]{c09} Cameron, P.B., Britton, M., \& 
Kulkarni, S. 2009, \aj, 137, 83
\bibitem[Cameron(2008)]{c08} Cameron, P.B. 2008, PhDT
\bibitem[Carpenter(2001)]{c01} Carpenter, J. 2001, \aj, 121, 2851
\bibitem[Chabrier(2001)]{ch01} Chabrier, G. 2001, \apj, 554, 1274
\bibitem[Chauvin et al.(2004)]{ch04} Chauvin, G., Lagrange, A., Dumas,
C., Zuckerman, B., Mouillet, D., Song, I., Beuzit, J., \& Lowrance, P. 
2004, \aap, 425, 29
\bibitem[Close et al.(2003)]{c03} Close, L., Siegler, N., Freed, M., \&
Biller, B. 2003, \apj, 587, 407
\bibitem[Close et al. (2007)]{c07} Close, L. et al. 2007, \apj, 660, 1492
\bibitem[Correia et al.(2006)]{cor06} Correia, S., Zinnecker, H., Ratzka, Th., 
\& Sterzik, M. 2006, \aap, 459, 909
\bibitem[de Zeeuw et al.(1999)]{dZ99} de Zeeuw, P., Hoogerwerf, R., de
Bruijne, J., Brown, A., \& Blaauw, A. 1999, \aj, 117, 354
\bibitem[Deacon \& Hambly(2007)]{dh07} Deacon, N. \& Hambly, N. 2007, \aap, 
468, 163
\bibitem[Duchene et al.(1999)]{d99} Duchene, G., Monin, J.-L., Bouvier, J., \& 
Menard, F. 1999, \aap, 351, 954
\bibitem[Duquennoy \& Mayor(1991)]{dm91} Duquennoy, A. \& Mayor, M. 1991, \aap, 248, 485
\bibitem[Durisen \& Sterzik(1994)]{ds94} Durisen, R. \& Sterzik, M. 1994, 
\aap, 286, 84
\bibitem[Elmegreen(2008)]{e08} Elmegreen, B. 2008, \apj, 672, 1006
\bibitem[Epchtein et al.(1999)]{ep99} Epchtein, N. et al. 1999, \aap, 349, 236
\bibitem[Fischer \& Marcy(1992)]{fm92} Fischer, D. \& Marcy, G. 1993, \apj, 396, 178
\bibitem[Frink et al. (1997)]{frink97} Frink, S., Roser, S, Neuhauser, R., \& Sterzik, M. 1997, 
\aap, 325, 613
\bibitem[Furlan et al.(2006)]{f06} Furlan, E. et al. 2006, \apjs, 165, 568
\bibitem[Ghez et al.(1993)]{g93} Ghez, A., Neugebauer, G., \& Matthews, K. 
1993, \aj, 106, 2005
\bibitem[Ghez et al.(2008)]{g08} Ghez, A. et al. 2008, \apj, 689, 1044
\bibitem[Gomez et al.(1993)]{gom93} Gomez, M., Hartmann, L., Kenyon, S., 
\& Hewett, R. 1993, \aj, 105, 1927
\bibitem[Guenther et al.(2007)]{guen07} Guenther, E., Esposito, M., Mundt, R., 
Covino, E., Alcala, J., Cusano, F., \& Stecklum, B. 2007, \aap, 467, 1147
\bibitem[Hambly et al.(2001)]{ham01} Hambly, N. et al. 2001, \mnras,
326, 1279
\bibitem[Hartigan et al.(1994)]{hart94} Hartigan, P., Strom, K., \& Strom, S. 
1994, \apj, 427, 961
\bibitem[Hartigan \& Kenyon(2003)]{hk03} Hartigan, P. \& Kenyon, S. 2003, \apj, 
583, 334
\bibitem[Hewett(1981)]{hew81} Hewett, P. 1982, \mnras, 201, 867
\bibitem[Hillenbrand(1997)]{hill97} Hillenbrand, L. 1997, \aj, 113, 1733
\bibitem[Hillenbrand \& Hartmann(1998)]{hh98} Hillenbrand, L. \& Hartmann, 
L. 1998, \apj, 492, 540
\bibitem[Hillenbrand \& White(2004)]{hw04} Hillenbrand, L. \& White, R.
2004, \apj, 604, 741
\bibitem[Itoh et al.(2005)]{itoh05} Itoh, Y. et al. 2005, \apj, 620, 984
\bibitem[Jayawardhana \& Ivanov(2006)]{ji06} Jayawardhana, R. \& Ivanov,
V. 2006, Science, 313, 1279
\bibitem[Jones \& Walker(1988)]{jw88} Jones, B. \& Walker, M. 1988, \aj, 
95, 1755
\bibitem[Kenyon \& Hartmann(1995)]{kh95} Kenyon, S. \& Hartmann, L. 1995, ApJS, 101, 117
\bibitem[Kim et al.(2005)]{kim05} Kim, S., Figer, D., Lee, M., \& Oh, S.
2005, \pasp, 117, 445
\bibitem[Kiminki et al.(2007)]{kim07} Kiminki, D. et al. 2007, \apj, 664, 1102
\bibitem[K\"ohler et al.(2000)]{koh00} K\"ohler, R. et al. 2000,
\aap, 356, 541
\bibitem[K\"ohler et al.(2006)]{koh06} K\"ohler, R., Petr-Gotzens, M., 
McCaughrean, M., Bouvier, J., Duch\^ene, G., Quirrenbach, A., \& 
Zinnecker, H. 2006, \aap, 458, 461
\bibitem[Kouwenhoven et al.(2007)]{kou07} Kouwenhoven, M.B.N., Brown, 
A.G.A., \& Kaper, L. 2007, \aap, 464, 581
\bibitem[Kouwenhoven et al.(2009)]{kou09} Kouwenhoven, M.B.N., 
Brown, A.G.A., Goodwin, S.P., Portegies Zwart, S.F., \& Kaper, L. 
2009, \aap, 493, 979
\bibitem[Kraus \& Hillenbrand(2007a)]{kraus07a} Kraus, A. \& Hillenbrand, A. 2007, \apj, 662, 413 (KH07a)
\bibitem[Kraus \& Hillenbrand(2007b)]{kraus07b} Kraus, A. \& Hillenbrand, A. 2007, \apj, 664, 1167
\bibitem[Kraus et al.(2008)]{kr08} Kraus, A., Ireland, M., Martinache, F., \& Lloyd, J. 2008, \apj, in press
\bibitem[Kroupa(1998)]{kro98} Kroupa, P. 1998, \mnras, 298, 231
\bibitem[Kroupa et al.(1999)]{kro99} Kroupa, P., Petr, M., \& McCaughrean, 
M. 1999, New Astronomy, 4, 495
\bibitem[Kroupa et al.(2001)]{kro01} Kroupa, P., Aarseth, S., \& 
Hurley, J. 2001, \mnras, 321, 699
\bibitem[Kroupa(2002)]{kro02} Kroupa, P. 2002, Science, 295, 82
\bibitem[Kroupa \& Bouvier(2003)]{kb03} Kroupa, P. \& Bouvier, J. 
2003, /mnras, 346, 343
\bibitem[Kunkel(1999)]{kun99} Kunkel, M. 1999, PhDT
\bibitem[Lada \& Lada(2003)]{ll03} Lada, J. \& Lada, E. 2003, ARA\&A, 41, 57
\bibitem[Larson(1995)]{lar95} Larson, R. 1995, \mnras, 272, 213
\bibitem[Leinert et al.(1993)]{lein93} Leinert, Ch. et al. 1993, \aap, 278, 129
\bibitem[Luhman \& Rieke(1998)]{lr98} Luhman, K. \& Rieke, G. 1998, \apj, 497, 354
\bibitem[Luhman(1999)]{luh99} Luhman, K. 1999, \apj, 525, 466
\bibitem[Luhman(2000)]{luh00} Luhman, K. 2000, \apj, 544, 1044
\bibitem[Luhman et al.(2003)]{luh03} Luhman, K., Stauffer, J., Muench,
Al, Rieke, G., Lada, E., Bouvier, J., \& Lada, C. 2003, \apj, 593, 1093
\bibitem[Luhman(2004)]{luh04} Luhman, K. 2004, \apj, 617, 1216
\bibitem[Luhman et al.(2006a)]{luh06a} Luhman, K. et al. 2006a, \apj, 649, 894
\bibitem[Luhman et al.(2006b)]{luh06b} Luhman, K., Whitney, B., Meade, M., 
Babler, B., Indebetouw, R., Bracker, S., \& Churchwell, E. 2006b, \apj, 647, 1180
\bibitem[Luhman(2006)]{luh06} Luhman, K. 2006, \apj, 645, 676
\bibitem[Luhman et al.(2007)]{luh07} Luhman, K., Allers, K., Jaffe, D.,
Cushing, M., Williams, K., Slesnick, C., \& Vacca, W. 2007, \apj, 659, 1629
\bibitem[MacConnell et al.(1992)]{mac92} MacConnell, D., Wing, R., \& Costa, E. 1992, \aj, 104, 821
\bibitem[Mart\'in et al.(2001)]{mar01} Mart\'in, E., Dougados, C., Magnier, E., 
Menard, F., Magazzu, A., Cuillandre, J., \& Delfosse, X. 2001, \apj, 561, 195
\bibitem[Mart\'in et al.(2004)]{mar04} Mart\'in, E., Delfosse, X., \& Guieu, S. 
2004, \aj, 127, 449
\bibitem[Massarotti et al.(2005)]{mas05} Massarotti, A., Latham, D., Torres, G., Brown, R., 
\& Oppenheimer, B. 2005,\aj, 129, 2294
\bibitem[Massey et al.(1988)]{mas88} Massey, P., Strobel, K., Barnes, J.,
\& Anderson, E. 1988, \apj, 328, 315
\bibitem[Metchev(2005)]{met05} Metchev, S. 2005, PhD 
thesis(http://etd.caltech.edu/etd/available/etd-08262005-160055/)
\bibitem[Miller \& Scalo(1979)]{ms79} Miller, G. \& Scalo, J. 1979, 
ApJS, 41, 513
\bibitem[Monet et al.(2003)]{monet03} Monet, D. et al. 2003, \aj, 125, 984
\bibitem[Oke \& Gunn(1982)]{oke82} Oke, B. \& Gunn, J. 1982, \pasp, 94, 586
\bibitem[Padgett et al.(2006)]{pad06} Padgett, D. et al. 2006, BAAS, 209, 3016
\bibitem[Patience et al.(2002)]{pat02} Patience, J. et al. 2002, \aj, 123, 1570
 \bibitem[Peebles(1980)]{peeb80} Peebles, J. 1980, The Large Scale Structure of 
the Universe (Princeton: Princeton Univ. Press)
\bibitem[Prato et al.(2002a)]{pr02a} Prato, L., Simon, M., Mazeh, T., McLean, 
I., Norman, D., \& Zucker, S. 2002a, \apj, 569, 863
\bibitem[Prato et al.(2002b)]{pr02b} Prato, L., Simon, M., Mazeh, T., Zucker, 
S., \& McLean, I. 2002b, \apj, 579, 99
\bibitem[Prato et al.(2007)]{pr07} Prato, L. 2007, \apj, 657, 338
\bibitem[Preibisch et al.(1998)]{pre98} Preibisch, T., Guenther, E., Zinnecker, 
H., Sterzik, M., Frink, S., \& Roeser, S. 1998, \aap, 333, 619
\bibitem[Preibisch et al.(2001)]{pre01} Preibisch, T. et al. 2001, \aj, 121, 1040
\bibitem[Preibisch et al.(2002)]{pre02} Preibisch, T. et al. 2002, \aj, 124, 404
\bibitem[Reid \& Gizis(1997)]{rg97} Reid, I.N. \& Gizis, J. 1997, \aj, 113, 2246
\bibitem[Reid et al.(2001)]{r01} Reid, I., Gizis, J., Kirkpatrick, J., \&
Koerner, D. 2001, \aj, 121, 489
\bibitem[Reipurth et al.(2007)]{rei07} Reipurth, B., Guimar\~aes, M., 
Connelley, M., \& Bally, J. 2007, \aj, 134, 2272
\bibitem[Salim \& Gould(2003)]{sg03} Salim, S. \& Gould, A. 2003, \apj, 582, 
1011
\bibitem[Scalo(1998)]{sca98} Scalo, J. 1998, in APS Conf. Ser. 142, 
The Stellar Initial Mass Function, ed. G. Gilmore \& D. Howell (San 
Francisco:ASP), 201
\bibitem[Scholz et al.(2006)]{sch06} Scholz, A., Jayawardhana, R., \& Wood, K. 2006, \apj, 645, 1498
\bibitem[Simon et al.(1995)]{sim95} Simon, M. et al. 1995, \apj, 443, 625
\bibitem[Simon et al.(1996)]{sim96} Simon, M., Holfeltz, S., Taff, L. 1996, 
\apj, 469, 890
\bibitem[Simon(1997)]{sim97} Simon, M. 1997, \apj, 482, 81
\bibitem[Simon et al.(2000)]{sim00} Simon, M., Dutrey, A., \& Guilloteau, S. 
2000, \apj, 545, 1034
\bibitem[Sirianni et al.(2002)]{sir02} Sirianni, M., Nota, A., De Marchi, 
G., Leitherer, C., \& Clampin, M. 2002, \apj, 579, 275
\bibitem[Skrutskie et al.(2006)]{skr06} Skrutskie, M. et al. 2006, \aj, 131, 1163
\bibitem[Slesnick et al.(2004)]{sles04} Slesnick, C., Hillenbrand, L., \& 
Carpenter, J. 2004, \apj, 610, 1045
\bibitem[Slesnick et al.(2006a)]{sles06a} Slesnick, C., Carpenter, J., \&
Hillenbrand, L. 2006a, \aj, 131, 3016
\bibitem[Slesnick et al.(2006b)]{sles06b} Slesnick, C., Carpenter, J.,
Hillenbrand, L., \& Mamajek, E. 2006b, \aj, 132, 2665
\bibitem[Stapelfeldt et al.(2003)]{sta03} Stapelfeldt, K., Menard, F., Watson, 
A., Krist, J., dougados, C., Padgett, D., \& Brandner, W. 2003, \apj, 589, 410
\bibitem[Steffen et al.(2001)]{stef01} Steffen, A. et al. 2001, \aj, 122, 997
\bibitem[Sterzik et al.(2003)]{ster03} Sterzik, M., Durisen, R., \& 
Zinnecker, H. 2003, \aap, 411, 91
\bibitem[Stetson(1987)]{stet87} Stetson, P. 1987, \pasp, 99, 191
\bibitem[Strom \& Strom(1994)]{ss94} Strom, K. \& Strom, S. 1994, \apj, 424, 237
\bibitem[Torres-Dodgen \& Weaver(1993)]{tdw93} Torres-Dodgen, A. \& Weaver, W.B. 1993, \pasp, 105, 693
\bibitem[Torres et al.(2007)]{tor07} Torres, R., Loinard, L., Mioduszewski, A., \& Rodriguez, L. 
2007, \apj, 671, 1813
\bibitem[Walter et al.(1994)]{wal94} Walter, F., Vrba, F., Mathieu, R., Brown, 
A., \& Myers, P. 1994, \aj, 107, 692
\bibitem[Weinberg et al.(1987)]{wein87} Weinberg, M., Shapiro, S., \&
Wasserman, I. 1987, \apj, 312, 367
\bibitem[White et al.(1999)]{w99} White, R., Ghez, A., Reid, I.N., \& Schultz, 
G. 1999, \apj, 520, 811
\bibitem[White \& Ghez(2001)]{wg01} White, R. \& Ghez, A. 2001, \apj, 556, 265
\bibitem[White \& Wing(1978)]{ww78} White, N. \& Wing, R. 1978, \apj, 222, 209
\bibitem[Wing(1971)]{w71} Wing, R. 1971, Late-Type Stars, Ed. G. W. Lockwood \& H. M. Dyck. Kitt 
Peak National Observatory, 554, 145
\bibitem[Wizinowich et al.(2006)]{wiz06} Wizinowich, P. et al. 2006, PASP,
118, 297
\bibitem[Zuckerman et al.(2004)]{zuck04} Zuckerman, B., Song, I., \& Bessell, M. 2004, \apj, 613, 65

\end{thebibliography}
\end{document}


\begin{deluxetable}{lllccl}
\tabletypesize{\scriptsize}
\tablewidth{0pt}
\tablecaption{Astrometric Data}
\tablehead{\colhead{Known Member} & \colhead{Candidate Companion} & 
\colhead{Epoch} & \colhead{Sep} & \colhead{PA} & \colhead{Ref}
\\
\colhead{} & \colhead{} & \colhead{(JD-2400000)} & \colhead{(mas)} & 
\colhead{(deg)}
}
\startdata
\multicolumn{6}{c}{New}\\
2M04080782+2807280&2M04080771+2807373&54069&9508$\pm$15&351.15$\pm$0.02&Keck-NGS\\
DG Tau&2M04270370+2606067&54434&16322$\pm$29&235.35$\pm$0.11&Palomar-NGS\\
GSC 06213-00306&GSC 06213-00306 B&54187&3213$\pm$2&306.3$\pm$0.02&Keck-NGS\\
GSC 06784-00997&2M16101888-2502325&54188&4896$\pm$2&241.24$\pm$0.02&Keck-NGS\\
GSC 06785-00476&2M15410726-2656254&54198&6270$\pm$10&82.65$\pm$0.1&Palomar-NGS\\
RXJ 1555.8-2512&2M15554839-2512174&54198&8877$\pm$14&319.73$\pm$0.1&Palomar-NGS\\
RXJ 1555.8-2512&2M15554788-2512172&54198&14524$\pm$23&299.27$\pm$0.1&Palomar-NGS\\
SCH161511.15-242015.56&2M16151239-2420091&54188&17885$\pm$22&70.24$\pm$0.07&Keck-LGS\\
SCH161825.01-233810.68&2M16182365-2338268&54199&24510$\pm$50&229.87$\pm$0.12&Palomar-Seeing\\
USco80&2M15583621-2348018&54188&12274$\pm$23&15.59$\pm$0.04&Keck-LGS\\
\multicolumn{6}{c}{Archival}\\
2M04080782+2807280&2M04080771+2807373&35403&7850$\pm$200&353.2$\pm$1.5&USNOB B1\\
2M04080782+2807280&2M04080771+2807373&48896&8620$\pm$200&351.0$\pm$1.3&USNOB B2\\
2M04080782+2807280&2M04080771+2807373&50781&9432$\pm$70&351.0$\pm$0.4&2MASS H\\
2M04080782+2807280&2M04080771+2807373&50781&9420$\pm$70&350.7$\pm$0.4&2MASS J\\
2M04080782+2807280&2M04080771+2807373&50781&9416$\pm$70&351.7$\pm$0.4&2MASS K\\
2M04161885+2752155&2M04161754+2751534&35403&28630$\pm$200&215.9$\pm$0.4&USNOB R1\\
2M04161885+2752155&2M04161754+2751534&47827&28000$\pm$200&218.3$\pm$0.4&USNOB R2\\
2M04161885+2752155&2M04161754+2751534&48896&27760$\pm$200&217.5$\pm$0.4&USNOB B2\\
2M04161885+2752155&2M04161754+2751534&50337&27970$\pm$200&218.7$\pm$0.4&USNOB I2\\
2M04161885+2752155&2M04161754+2751534&50782&28063$\pm$70&218.3$\pm$0.1&2MASS H\\
2M04161885+2752155&2M04161754+2751534&50782&28033$\pm$70&218.3$\pm$0.1&2MASS J\\
2M04213460+2701388&2M04213331+2701375&33626&17540$\pm$200&263.3$\pm$0.7&USNOB R1\\
2M04213460+2701388&2M04213331+2701375&47888&16970$\pm$200&265.6$\pm$0.7&USNOB B2\\
2M04213460+2701388&2M04213331+2701375&49636&17460$\pm$200&266.1$\pm$0.7&USNOB R2\\
2M04213460+2701388&2M04213331+2701375&50071&17400$\pm$200&265.6$\pm$0.7&USNOB I2\\
2M04213460+2701388&2M04213331+2701375&50782&17317$\pm$70&265.6$\pm$0.2&2MASS H\\
2M04213460+2701388&2M04213331+2701375&50782&17051$\pm$70&265.7$\pm$0.2&2MASS J\\
2M04213460+2701388&2M04213331+2701375&50782&17124$\pm$70&266.2$\pm$0.2&2MASS K\\
CFHT-Tau-21&2M04221757+2654364&33626&23760$\pm$200&151.0$\pm$0.5&USNOB B1\\
CFHT-Tau-21&2M04221757+2654364&33626&23700$\pm$200&151.5$\pm$0.5&USNOB R1\\
CFHT-Tau-21&2M04221757+2654364&47888&23360$\pm$200&152.2$\pm$0.5&USNOB B2\\
CFHT-Tau-21&2M04221757+2654364&49636&22950$\pm$200&151.8$\pm$0.5&USNOB R2\\
CFHT-Tau-21&2M04221757+2654364&50071&23320$\pm$200&152.3$\pm$0.5&USNOB I2\\
CFHT-Tau-21&2M04221757+2654364&50782&23315$\pm$70&152.1$\pm$0.2&2MASS H\\
CFHT-Tau-21&2M04221757+2654364&50782&23327$\pm$70&152.0$\pm$0.2&2MASS J\\
CFHT-Tau-21&2M04221757+2654364&50782&23244$\pm$70&152.3$\pm$0.2&2MASS K\\
CFHT-Tau-7&2M04321713+2421556&33626&22890$\pm$300&206.0$\pm$0.8&USNOB B1\\
CFHT-Tau-7&2M04321713+2421556&33626&22340$\pm$200&205.6$\pm$0.5&USNOB R1\\
CFHT-Tau-7&2M04321713+2421556&47888&22190$\pm$200&206.8$\pm$0.5&USNOB B2\\
CFHT-Tau-7&2M04321713+2421556&49636&22230$\pm$200&206.3$\pm$0.5&USNOB R2\\
CFHT-Tau-7&2M04321713+2421556&50071&21910$\pm$200&206.7$\pm$0.5&USNOB I2\\
CFHT-Tau-7&2M04321713+2421556&50783&21770$\pm$70&207.2$\pm$0.2&2MASS H\\
CFHT-Tau-7&2M04321713+2421556&50783&21797$\pm$70&207.2$\pm$0.2&2MASS J\\
CFHT-Tau-7&2M04321713+2421556&50783&21736$\pm$70&207.4$\pm$0.2&2MASS K\\
DG Tau&2M04270370+2606067&50783&16494$\pm$70&234.3$\pm$0.2&2MASS H\\
DG Tau&2M04270370+2606067&50783&16296$\pm$70&234.7$\pm$0.2&2MASS J\\
DG Tau&2M04270370+2606067&50783&16597$\pm$70&235.3$\pm$0.2&2MASS K\\
FO Tau&2M04144741+2812219&35403&26270$\pm$200&240.3$\pm$0.4&USNOB R1\\
FO Tau&2M04144741+2812219&47827&26100$\pm$200&249.7$\pm$0.4&USNOB R2\\
FO Tau&2M04144741+2812219&50337&26460$\pm$200&250.3$\pm$0.4&USNOB I2\\
FO Tau&2M04144741+2812219&50781&26051$\pm$70&249.9$\pm$0.2&2MASS H\\
FO Tau&2M04144741+2812219&50781&26152$\pm$70&250.5$\pm$0.2&2MASS J\\
FO Tau&2M04144741+2812219&50781&26461$\pm$70&250.8$\pm$0.2&2MASS K\\
FW Tau&2M04292887+2616483&33626&12740$\pm$200&241.4$\pm$0.9&USNOB B1\\
FW Tau&2M04292887+2616483&33626&12720$\pm$200&241.1$\pm$0.9&USNOB R1\\
FW Tau&2M04292887+2616483&47888&12260$\pm$200&245.9$\pm$0.9&USNOB B2\\
FW Tau&2M04292887+2616483&49636&12340$\pm$200&246.3$\pm$0.9&USNOB R2\\
FW Tau&2M04292887+2616483&50783&12246$\pm$70&247.0$\pm$0.3&2MASS H\\
FW Tau&2M04292887+2616483&50783&12200$\pm$70&246.7$\pm$0.3&2MASS J\\
FW Tau&2M04292887+2616483&50783&12229$\pm$70&246.6$\pm$0.3&2MASS K\\
GM Aur&2M04551015+3021333&35106&29470$\pm$200&200.9$\pm$0.4&USNOB B1\\
GM Aur&2M04551015+3021333&35106&29090$\pm$200&201.4$\pm$0.4&USNOB R1\\
GM Aur&2M04551015+3021333&50458&28360$\pm$200&202.2$\pm$0.4&USNOB I2\\
GM Aur&2M04551015+3021333&50844&28301$\pm$70&202.1$\pm$0.1&2MASS H\\
GM Aur&2M04551015+3021333&50844&28289$\pm$70&202.1$\pm$0.1&2MASS J\\
GM Aur&2M04551015+3021333&50844&28295$\pm$70&202.1$\pm$0.1&2MASS K\\
GM Aur&2M04551015+3021333&51462&28370$\pm$200&202.0$\pm$0.4&USNOB R2\\
HBC 427&2M04560252+3020503&35106&10440$\pm$200&158.6$\pm$1.1&USNOB R1\\
HBC 427&2M04560252+3020503&50101&12010$\pm$200&155.5$\pm$1&USNOB B2\\
HBC 427&2M04560252+3020503&50458&12330$\pm$200&156.8$\pm$0.9&USNOB I2\\
HBC 427&2M04560252+3020503&50844&14930$\pm$70&154.5$\pm$0.3&2MASS H\\
HBC 427&2M04560252+3020503&50844&14934$\pm$70&154.6$\pm$0.3&2MASS J\\
HBC 427&2M04560252+3020503&50844&14973$\pm$70&154.3$\pm$0.3&2MASS K\\
HBC 427&2M04560252+3020503&51462&11970$\pm$200&155.7$\pm$1&USNOB R2\\
I04385+2550&2M04413842+2556448&33626&18030$\pm$200&341.2$\pm$0.6&USNOB R1\\
I04385+2550&2M04413842+2556448&47892&18970$\pm$200&343.1$\pm$0.6&USNOB B2\\
I04385+2550&2M04413842+2556448&48296&18240$\pm$200&341.8$\pm$0.6&USNOB R2\\
I04385+2550&2M04413842+2556448&50336&18890$\pm$200&342.4$\pm$0.6&USNOB I2\\
I04385+2550&2M04413842+2556448&51101&18893$\pm$70&343.4$\pm$0.2&2MASS H\\
I04385+2550&2M04413842+2556448&51101&18901$\pm$70&343.1$\pm$0.2&2MASS J\\
I04385+2550&2M04413842+2556448&51101&18971$\pm$70&343.2$\pm$0.2&2MASS K\\
IS Tau&2M04333467+2609447&33626&28530$\pm$200&258.4$\pm$0.4&USNOB R1\\
IS Tau&2M04333467+2609447&49636&28950$\pm$200&261.3$\pm$0.4&USNOB R2\\
IS Tau&2M04333467+2609447&50071&28590$\pm$200&261.0$\pm$0.4&USNOB I2\\
IS Tau&2M04333467+2609447&50783&28649$\pm$70&261.7$\pm$0.1&2MASS H\\
IS Tau&2M04333467+2609447&50783&28781$\pm$70&261.7$\pm$0.1&2MASS J\\
IS Tau&2M04333467+2609447&50783&28724$\pm$70&261.6$\pm$0.1&2MASS K\\
ScoPMS048&ScoPMS 048 B&49910&3429$\pm$9&192.3$\pm$0.2&K\"ohler et al. (2000)\\
ScoPMS048&ScoPMS 048 B&50693&3410$\pm$10&192.6$\pm$0.2&Massarotti et al. (2005)\\
ScoPMS048&ScoPMS 048 B&51297&3335$\pm$70&190.9$\pm$1.2&2MASS H\\
ScoPMS048&ScoPMS 048 B&51297&3387$\pm$70&192.5$\pm$1.2&2MASS J\\
ScoPMS048&ScoPMS 048 B&51297&3509$\pm$70&191.7$\pm$1.1&2MASS K\\
ScoPMS048&ScoPMS 048 B&54252&3419$\pm$6&192.58$\pm$0.10&Kraus et al. (2008)\\
V710 Tau AB&2M04315968+1821305&35452&27580$\pm$200&102.7$\pm$0.4&USNOB B1\\
V710 Tau AB&2M04315968+1821305&35452&27720$\pm$200&102.8$\pm$0.4&USNOB R1\\
V710 Tau AB&2M04315968+1821305&48541&27740$\pm$200&102.4$\pm$0.4&USNOB R2\\
V710 Tau AB&2M04315968+1821305&48653&27440$\pm$200&102.2$\pm$0.4&USNOB B2\\
V710 Tau AB&2M04315968+1821305&50051&27780$\pm$200&101.9$\pm$0.4&USNOB I2\\
V710 Tau AB&2M04315968+1821305&51872&27481$\pm$70&101.8$\pm$0.1&2MASS H\\
V710 Tau AB&2M04315968+1821305&51872&27604$\pm$70&102.1$\pm$0.1&2MASS J\\
V710 Tau AB&2M04315968+1821305&51872&27508$\pm$70&102.3$\pm$0.1&2MASS K\\
GSC 06213-00306&GSC 06213-00306 B&51297&3161$\pm$70&305.4$\pm$1.3&2MASS H\\
GSC 06213-00306&GSC 06213-00306 B&51297&3176$\pm$70&306.1$\pm$1.3&2MASS J\\
GSC 06213-00306&GSC 06213-00306 B&51297&3193$\pm$70&305.1$\pm$1.3&2MASS K\\
GSC 06213-00306&GSC 06213-00306 B&54252&3210$\pm$5&306.46$\pm$0.10&Kraus et al. (2008)\\
GSC 06784-00997&2M16101888-2502325&50998&4899$\pm$70&240.6$\pm$0.8&2MASS H\\
GSC 06784-00997&2M16101888-2502325&50998&4943$\pm$70&240.8$\pm$0.8&2MASS J\\
GSC 06784-00997&2M16101888-2502325&50998&4884$\pm$70&238.9$\pm$0.8&2MASS K\\
GSC 06785-00476&2M15410726-2656254&51305&6267$\pm$70&82.0$\pm$0.6&2MASS H\\
GSC 06785-00476&2M15410726-2656254&51305&6327$\pm$70&82.0$\pm$0.6&2MASS J\\
GSC 06785-00476&2M15410726-2656254&51305&6328$\pm$70&82.4$\pm$0.6&2MASS K\\
GSC 06785-00476&2M15410726-2656254&52836&6261$\pm$18&82.1$\pm$0.2&Metchev (2005)\\
RXJ 1555.8-2512&2M15554788-2512172&51306&14620$\pm$70&298.5$\pm$0.3&2MASS H\\
RXJ 1555.8-2512&2M15554839-2512174&51306&8853$\pm$70&319.3$\pm$0.5&2MASS H\\
RXJ 1555.8-2512&2M15554788-2512172&51306&14563$\pm$70&298.4$\pm$0.3&2MASS J\\
RXJ 1555.8-2512&2M15554839-2512174&51306&8842$\pm$70&319.0$\pm$0.5&2MASS J\\
RXJ 1555.8-2512&2M15554788-2512172&51306&14647$\pm$70&298.3$\pm$0.3&2MASS K\\
RXJ 1555.8-2512&2M15554839-2512174&51306&8846$\pm$70&319.1$\pm$0.5&2MASS K\\
RXJ 1555.8-2512&2M15554839-2512174&51662&8830$\pm$70&319.2$\pm$0.5&DENIS\\
RXJ 1555.8-2512&2M15554788-2512172&51662&14398$\pm$70&299.1$\pm$0.3&DENIS\\
RXJ 1558.8-2512&2M15585415-2512407&34834&11820$\pm$200&134.8$\pm$1&USNOB B1\\
RXJ 1558.8-2512&2M15585415-2512407&34834&11290$\pm$200&134.1$\pm$1&USNOB R1\\
RXJ 1558.8-2512&2M15585415-2512407&42925&11390$\pm$200&132.8$\pm$1&USNOB B2\\
RXJ 1558.8-2512&2M15585415-2512407&50997&11383$\pm$70&130.1$\pm$0.4&2MASS H\\
RXJ 1558.8-2512&2M15585415-2512407&50997&11308$\pm$70&130.1$\pm$0.4&2MASS J\\
RXJ 1558.8-2512&2M15585415-2512407&50997&11457$\pm$70&129.6$\pm$0.4&2MASS K\\
RXJ 1558.8-2512&2M15585415-2512407&51759&11230$\pm$70&129.8$\pm$0.4&DENIS\\
RXJ 1602.8-2401B&2M16025116-2401502&50215&6990$\pm$70&351.5$\pm$0.6&DENIS\\
RXJ 1602.8-2401B&2M16025116-2401502&50998&7262$\pm$70&353.5$\pm$0.6&2MASS H\\
RXJ 1602.8-2401B&2M16025116-2401502&50998&7230$\pm$70&353.2$\pm$0.6&2MASS J\\
RXJ 1602.8-2401B&2M16025116-2401502&50998&7214$\pm$70&351.6$\pm$0.6&2MASS K\\
RXJ 1602.8-2401B&2M16025116-2401502&54252&7253$\pm$14&353.56$\pm$0.11&Kraus et al. (2008)\\
SCH160758.50-203948.90&2M16075693-2039424&33447&23080$\pm$200&282.5$\pm$0.5&USNOB R1\\
SCH160758.50-203948.90&2M16075693-2039424&42958&24290$\pm$200&283.6$\pm$0.5&USNOB B2\\
SCH160758.50-203948.90&2M16075693-2039424&46244&22870$\pm$200&284.9$\pm$0.5&USNOB I2\\
SCH160758.50-203948.90&2M16075693-2039424&48338&23880$\pm$200&284.7$\pm$0.5&USNOB R2\\
SCH160758.50-203948.90&2M16075693-2039424&50932&22990$\pm$70&285.4$\pm$0.2&DENIS\\
SCH160758.50-203948.90&2M16075693-2039424&51297&23240$\pm$70&285.1$\pm$0.2&2MASS H\\
SCH160758.50-203948.90&2M16075693-2039424&51297&22816$\pm$70&285.6$\pm$0.2&2MASS J\\
SCH160758.50-203948.90&2M16075693-2039424&51297&22704$\pm$70&285.4$\pm$0.2&2MASS K\\
SCH161511.15-242015.56&2M16151239-2420091&49479&17950$\pm$200&69.2$\pm$0.6&USNOB I2\\
SCH161511.15-242015.56&2M16151239-2420091&50218&17810$\pm$70&70.0$\pm$0.2&DENIS\\
SCH161511.15-242015.56&2M16151239-2420091&50998&18002$\pm$70&69.7$\pm$0.2&2MASS H\\
SCH161511.15-242015.56&2M16151239-2420091&50998&17896$\pm$70&69.6$\pm$0.2&2MASS J\\
SCH161511.15-242015.56&2M16151239-2420091&50998&18037$\pm$70&70.0$\pm$0.2&2MASS K\\
SCH161825.01-233810.68&2M16182365-2338268&42928&24810$\pm$200&227.9$\pm$0.5&USNOB B2\\
SCH161825.01-233810.68&2M16182365-2338268&48340&25260$\pm$200&228.1$\pm$0.5&USNOB R2\\
SCH161825.01-233810.68&2M16182365-2338268&49479&25150$\pm$200&228.4$\pm$0.5&USNOB I2\\
SCH161825.01-233810.68&2M16182365-2338268&51297&24823$\pm$70&229.1$\pm$0.2&2MASS H\\
SCH161825.01-233810.68&2M16182365-2338268&51297&24695$\pm$70&229.0$\pm$0.2&2MASS J\\
SCH161825.01-233810.68&2M16182365-2338268&51297&24783$\pm$70&229.0$\pm$0.2&2MASS K\\
SCH161825.01-233810.68&2M16182365-2338268&51341&24760$\pm$70&229.1$\pm$0.2&DENIS\\
SCH162135.91-235503.41&2M16213638-2355283&45180&26300$\pm$200&165.5$\pm$0.4&USNOB I2\\
SCH162135.91-235503.41&2M16213638-2355283&49076&26080$\pm$200&165.0$\pm$0.4&USNOB R2\\
SCH162135.91-235503.41&2M16213638-2355283&50200&25760$\pm$70&165.4$\pm$0.2&DENIS\\
SCH162135.91-235503.41&2M16213638-2355283&50547&25630$\pm$70&165.5$\pm$0.2&DENIS\\
SCH162135.91-235503.41&2M16213638-2355283&51297&25644$\pm$70&165.2$\pm$0.2&2MASS H\\
SCH162135.91-235503.41&2M16213638-2355283&51297&25631$\pm$70&165.1$\pm$0.2&2MASS J\\
SCH162135.91-235503.41&2M16213638-2355283&51297&25440$\pm$70&164.9$\pm$0.2&2MASS K\\
ScoPMS042b&2M16102177-1904021&50140&4604$\pm$11&6.5$\pm$0.1&K\"ohler et al. (2000)\\
ScoPMS042b&2M16102177-1904021&51297&4623$\pm$70&6.5$\pm$0.9&2MASS H\\
ScoPMS042b&2M16102177-1904021&51297&4575$\pm$70&5.8$\pm$0.9&2MASS J\\
ScoPMS042b&2M16102177-1904021&51297&4578$\pm$70&6.0$\pm$0.9&2MASS K\\
ScoPMS042b&2M16102177-1904021&54256&4612$\pm$2&6.84$\pm$0.02&Kraus et al. (2008)\\
USco80&2M15583621-2348018&34834&12270$\pm$200&16.0$\pm$0.9&USNOB R1\\
USco80&2M15583621-2348018&42925&12690$\pm$200&16.8$\pm$0.9&USNOB B2\\
USco80&2M15583621-2348018&48340&12140$\pm$200&14.5$\pm$0.9&USNOB R2\\
USco80&2M15583621-2348018&49479&11590$\pm$200&15.9$\pm$1&USNOB I2\\
USco80&2M15583621-2348018&51296&12288$\pm$70&15.0$\pm$0.3&2MASS H\\
USco80&2M15583621-2348018&51296&12239$\pm$70&14.9$\pm$0.3&2MASS J\\
USco80&2M15583621-2348018&51296&12253$\pm$70&15.2$\pm$0.3&2MASS K\\
USco80&2M15583621-2348018&51759&12230$\pm$70&15.2$\pm$0.3&DENIS\\
UScoJ160245.4-193037&2M16024735-1930294&33447&25800$\pm$200&71.2$\pm$0.4&USNOB R1\\
UScoJ160245.4-193037&2M16024735-1930294&42958&26730$\pm$200&71.7$\pm$0.4&USNOB B2\\
UScoJ160245.4-193037&2M16024735-1930294&46244&27450$\pm$200&72.3$\pm$0.4&USNOB I2\\
UScoJ160245.4-193037&2M16024735-1930294&48338&27450$\pm$200&72.6$\pm$0.4&USNOB R2\\
UScoJ160245.4-193037&2M16024735-1930294&50215&27780$\pm$70&72.8$\pm$0.1&DENIS\\
UScoJ160245.4-193037&2M16024735-1930294&50990&28207$\pm$70&73.0$\pm$0.1&2MASS H\\
UScoJ160245.4-193037&2M16024735-1930294&50990&28179$\pm$70&72.9$\pm$0.1&2MASS J\\
UScoJ160245.4-193037&2M16024735-1930294&50990&28128$\pm$70&72.9$\pm$0.1&2MASS K\\
UScoJ160700.1-203309&2M16065937-2033047&33447&11780$\pm$200&292.9$\pm$1&USNOB B1\\
UScoJ160700.1-203309&2M16065937-2033047&33447&11400$\pm$200&292.7$\pm$1&USNOB R1\\
UScoJ160700.1-203309&2M16065937-2033047&42958&11990$\pm$200&291.9$\pm$1&USNOB B2\\
UScoJ160700.1-203309&2M16065937-2033047&46244&11020$\pm$200&292.9$\pm$1&USNOB I2\\
UScoJ160700.1-203309&2M16065937-2033047&48338&11120$\pm$200&291.4$\pm$1&USNOB R2\\
UScoJ160700.1-203309&2M16065937-2033047&50174&11750$\pm$70&293.0$\pm$0.3&DENIS\\
UScoJ160700.1-203309&2M16065937-2033047&51297&11787$\pm$70&292.8$\pm$0.3&2MASS H\\
UScoJ160700.1-203309&2M16065937-2033047&51297&11829$\pm$70&292.4$\pm$0.3&2MASS J\\
UScoJ160700.1-203309&2M16065937-2033047&51297&11806$\pm$70&292.8$\pm$0.3&2MASS K\\
UScoJ160700.1-203309&2M16065937-2033047&51688&11800$\pm$70&292.8$\pm$0.3&DENIS\\
UScoJ160936.5-184800&2M16093658-1847409&33447&19740$\pm$200&358.2$\pm$0.6&USNOB B1\\
UScoJ160936.5-184800&2M16093658-1847409&33447&19140$\pm$200&357.9$\pm$0.6&USNOB R1\\
UScoJ160936.5-184800&2M16093658-1847409&42958&19510$\pm$200&0.5$\pm$0.6&USNOB B2\\
UScoJ160936.5-184800&2M16093658-1847409&46244&19350$\pm$200&1.1$\pm$0.6&USNOB I2\\
UScoJ160936.5-184800&2M16093658-1847409&48338&19740$\pm$200&1.4$\pm$0.6&USNOB R2\\
UScoJ160936.5-184800&2M16093658-1847409&50216&19950$\pm$70&2.0$\pm$0.2&DENIS\\
UScoJ160936.5-184800&2M16093658-1847409&51115&20040$\pm$70&2.6$\pm$0.2&DENIS\\
UScoJ160936.5-184800&2M16093658-1847409&51297&19934$\pm$70&2.3$\pm$0.2&2MASS H\\
UScoJ160936.5-184800&2M16093658-1847409&51297&19980$\pm$70&2.2$\pm$0.2&2MASS J\\
UScoJ160936.5-184800&2M16093658-1847409&51297&19913$\pm$70&1.9$\pm$0.2&2MASS K\\
UScoJ161031.9-191305&2M16103232-1913085&51297&5872$\pm$70&112.6$\pm$0.7&2MASS H\\
UScoJ161031.9-191305&2M16103232-1913085&51297&5762$\pm$70&114.1$\pm$0.7&2MASS J\\
UScoJ161031.9-191305&2M16103232-1913085&51297&5965$\pm$70&113.6$\pm$0.7&2MASS K\\
UScoJ161031.9-191305&2M16103232-1913085&54253&5820$\pm$9&114.01$\pm$0.10&Kraus et al. (2008)\\
\enddata
\end{deluxetable}